\documentclass[aps,prd,twocolumn,showkeys,groupedaddress,nofootinbib]{revtex4-1}
\usepackage{graphicx}
\usepackage{subcaption}
\usepackage{filecontents}
\usepackage{bm}
\usepackage{hyperref}
\usepackage{cleveref}

\usepackage{amsmath,amssymb,amsfonts,amsbsy}
\usepackage{epsfig}
\usepackage{color}                                                       %
\usepackage{latexsym} 
\usepackage{mathrsfs} 
\usepackage{slashed} 
\usepackage{psfrag}

\def\RE{{\rm Re}} 
\def\IM{{\rm Im}}
\def\Tr{\mathop{\rm Tr}\nolimits}
\def\vecsigma{\boldsymbol{\sigma}}

\def\Pcal{{\bf{\cal P}}}

\def\vers{{\cal V}}
\def\pros{{\cal D}}
\def\g{\Delta}

\def\un{\mathbf{1}} 

\def\sv{{\bf{s}}}

\def\pv{{\bf{p}}}
\def\qv{{\bf{q}}}

\def\rv{{\bf{r}}}
\def\kv{{\bf{k}}}

\def\Hv{{\bf H}}
\def\Pv{{\bf{P}}}
\def\Sv{{\bf{S}}}

\def\Lv{{\bf L}}

\def\Pv{{\bf P}}

\def\xu{\hat{\bf{x}}}
\def\yu{\hat{\bf{y}}}
\def\zu{\hat{\bf{z}}}
\def\ntil{\tilde{\bf{n}}}

\def\Mcal{{\mathcal{M}}}

\def\Pcal{{\cal P}}

\def\barr{\begin{eqnarray}}
\def\earr{\end{eqnarray}}
\def\ni{\noindent}

\def\ie{{\sl i.e.}}

\def\kt{     { \bf{k}_{\rm T}  }    }
\def\qt{     { \bf{k}_{\rm T}  }    }
\def\kpt{  \kv'_{\rm T} }
\def\ktkt{\kv^2_{\rm T}}
\def\kptkpt{{\kv'}^2_{\rm T}}

\def\pt{     { \bf{p}_{\rm T}  }    }
\def\ptpt{\pv^2_{\rm T}}
\def\bt {b\T}
\def\bl {b\L}
\def\T{_{\rm T}}
\def\L{_{\rm L}}
\def\A{_{\rm A}}
\def\B{_{\rm B}}
\def\q{\mathfrak{q}}

\def\QA{q_{\rm A}}

\def\QbB{\bar{q}_{\rm B}}
\def\h{\mathfrak{h}}
\def\A{_{\rm A}}
\def\B{_{\rm B}}
 
\def\R{{\bf Q}}

\def\be{\begin{equation}}
\def\ee{\end{equation}}

\begin{document}


\title{Recursive model for the fragmentation of polarized quarks}

\author{ A. Kerbizi$^{\, 1}$, X. Artru$^{\, 2}$, Z. Belghobsi$^{\, 3}$, F. Bradamante$^{\, 1}$ and A. Martin$^{\, 1}$}

\affiliation{$^{1}$ {\small INFN Sezione di Trieste and Dipartimento di Fisica, Universit\`a di Trieste,}\\
{\small Via Valerio 2, 34127 Trieste, Italy}\\
$^{2}$ {\small Univ. Lyon, Universit\'e Lyon 1, CNRS,}
{\small Institut de Physique Nucl\'eaire de Lyon, 69622 Villeurbanne, France}
\\$^{3}$ {\small Laboratoire de Physique Th\'eorique, Facult\'e des Sciences Exactes et de l'Informatique,}\\ 
{\small Universit\'e Mohammed Seddik Ben Yahia,}\\
{\small B.P. 98 Ouled Aissa, 18000 Jijel, Algeria}\\
}


\begin{abstract}
We present a model for Monte Carlo simulation of the fragmentation of a polarized quark. The model is based on string dynamics and the ${}^3P_0$ mechanism of quark pair creation at string breaking. 
The fragmentation is treated as a recursive process, where the splitting function of the subprocess $q \to h + q'$ depends on the spin density matrix of the quark $q$. The ${}^3P_0$ mechanism is parametrized by a complex mass parameter $\mu$, the imaginary part of which is responsible for single spin asymmetries. The model has been implemented in a Monte Carlo program to simulate jets made of pseudoscalar mesons. Results for single hadron and hadron pair transverse-spin asymmetries are found to be in agreement with experimental data from SIDIS and $e^+e^-$ annihilation. The model predictions on the jet-handedness are also discussed.

\end{abstract}

\date{\today}


\maketitle

\section{Introduction}
The fragmentation of quarks into hadron jets is a non perturbative QCD process and as such still poorly understood theoretically. For this reason many simulation models based on the recursive splitting process $q\to h+q'$ 
\cite{feynman-fieldw, Fi-Pe}  have been developped to account for the main jet properties, prepare high-energy experiments and analyze their results. The most elaborate one is the Lund Symmetric Model (LSM) \cite{lund}, based on the semiclassical dynamics of a string or color flux tube. This model is the basis of the fragmentation part in the event generator PYTHIA \cite{pythia}, which successfully describes the experimental data for different scattering processes. 
Up to now, however, the available codes of quark fragmentation do not include the quark spin degree of freedom, at least in a systematic way, like they do for flavor.
In particular they cannot simulate the Collins effect and the di-hadron transverse-spin asymmetries observed in semi-inclusive deep inelastic scattering (SIDIS) and in $e^+e^-$ annihilation. 
Progresses toward the inclusion of the quark spin has been done in Refs. \cite{DS09,DS11,DS13},  
where it was proposed to extend the LSM by assuming that the quark-antiquark pairs created during string breaking 
are in the $^3P_0$ state, which means that their spins are parallel and they have one unit of orbital angular momentum, antiparallel to the
total spin. The model of Refs. \cite{DS09,DS11,DS13} is a quantum version of the classical \textit{string + $^3P_0$} mechanism proposed in order to explain the inclusive hyperon polarization \cite{lund} and used to explain the single-spin asymmetry in $p\uparrow + p \to \pi + X$ \cite{XA-JCZ}.

The road map proposed in Refs. \cite{DS09,DS11,DS13} has now been pursued, and in this paper we give full account of the model which has been developed and of the program we have written to simulate the fragmentation of a polarized quark. The main results obtained will also be given, as well as their comparison with the data.  Preliminary results were presented in \cite{albi-tesi,spin16}.

The paper is organised as follows.
The theoretical framework of the recursive model on which the code is based is summarized
in Section II. The details of the string model, first in the spinless case, and then
introducing the spin and the $^3P_0$ model, are spelled out in Section III. The implementation
of the model in a Monte Carlo code is the subject of Section IV, while the results of the
simulation are compared with corresponding data from the COMPASS and BELLE Collaborations in Section V.
Section VI is dedicated to the simulation of jet-handness, and our conclusions are drawn
in Section VII.

For completeness, we remind that a different
model of polarized quark fragmentation, based  on  the  Nambu - Jona  Lasinio quark-meson coupling, has recently been proposed in Ref. \cite{Mate}.


\section{The general recursive model}
\begin{figure}[tb]
\begin{minipage}{.4\textwidth}
     \includegraphics[width=1.2\linewidth]{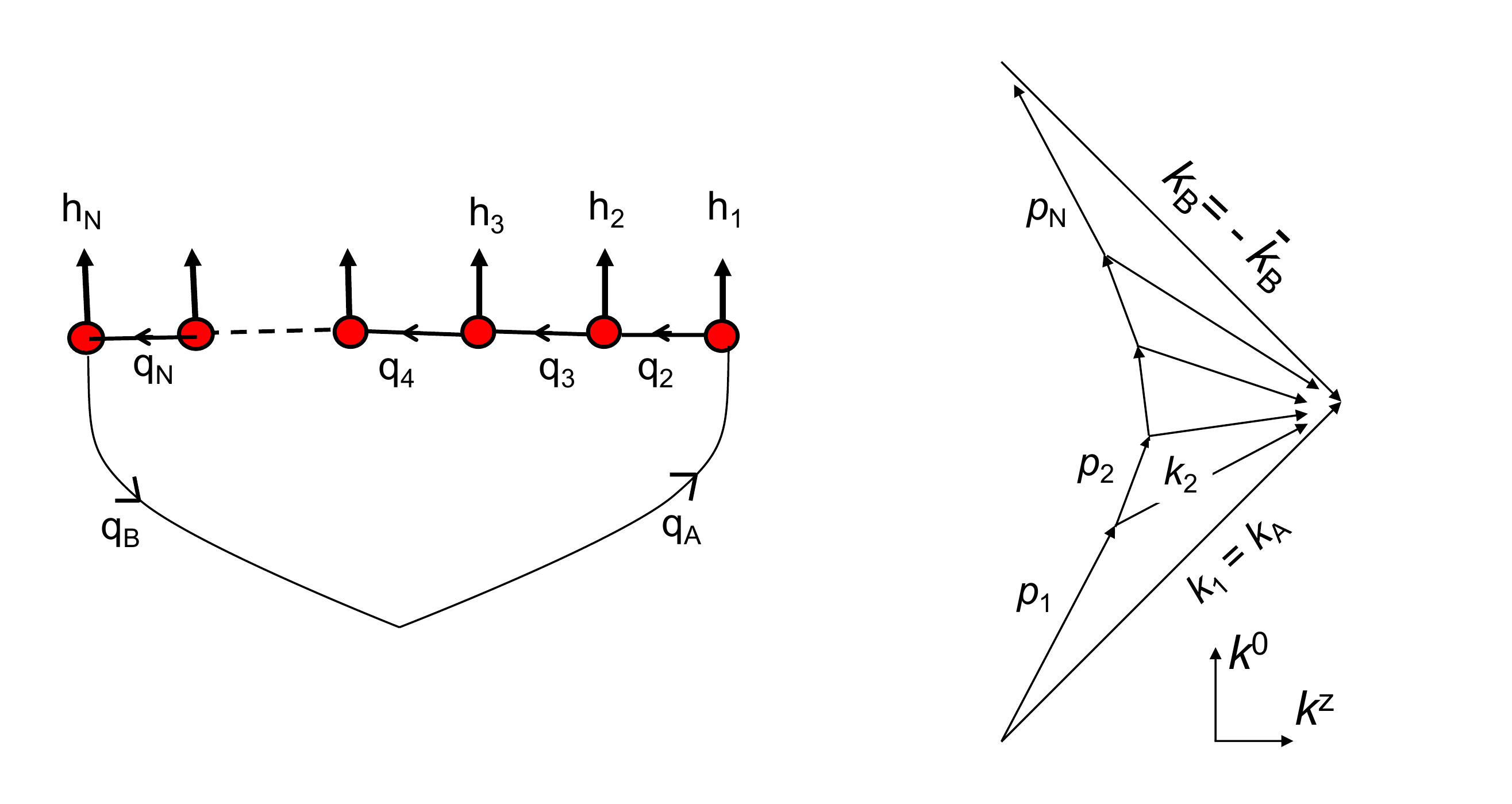}
     \end{minipage}
 \caption{\small{Left: multiperipheral diagram. Right: the associated momentum diagram.}}
  \label{fig:multiperipheral}
\end{figure}

We consider the hadronization  
\be \label{hadronization}
q_A+\bar{q}_B\rightarrow h_1+h_2+\dots + h_N
\ee 
occuring in $e^+e^-$ annihilation or $W^\pm$ decay in two jets. In SIDIS the antiquark $\bar{q}_B$ is replaced by the target remnant represented by a diquark. This process is  supposed to occur via a quark chain diagram shown in Fig. \ref{fig:multiperipheral} and modelised as the set of splittings

\begin{eqnarray} \label{recursif}
q\A\rightarrow h_1+q_2, \, ~~ q_2\rightarrow h_2 +q_{3}, \, \dots ~~~
\nonumber\\
\dots \, q_r\rightarrow h_r +q_{r+1}, \, \dots\, q_N\rightarrow h_N +q\B
\end{eqnarray}   
\ie, as the iteration of the elementary splitting
\begin{equation}
\label{eq: splitt elementary} 
q\rightarrow h + q'
\end{equation}
where the flavour content of the hadron $h$ is $q\bar{q}'$. 
The index $r = 1, 2 \dots N$ in Eq. (\ref{recursif}) is the \textit{rank} of the hadron or of the splitting quark.
The production of baryons is not included in the present code. 
In the following $k$ denotes the 4-momentum of a quark, $p$ that of a hadron.  In Eq. (\ref{eq: splitt elementary}) momentum conservation gives $p=k-k'$.
In the recursive model one assumes that the initial 4-momenta $k_1\equiv k\A$ and $k_{\overline{\rm B}}\equiv - k\B$ are on mass shell and generated beforehand\footnote{This is a classical approximation: considering Fig. \ref{fig:multiperipheral} as a loop diagram, $k\A$ is an internal momentum.  The cross section is then of the form $\int d^4k\A\, A(k\A, \cdots) \int d^4k'\A \, A^*(k'\A, \cdots)$ with $k\A$ and $k'\A$ being generally different and off mass shell.}.
In the $q\A \bar q\B$ centre-of-mass frame we orient the $\zu$ axis (named \textit{jet axis}) along  $\kv\A$. 
In SIDIS this axis usually differs from that defined in the laboratory frame by the virtual photon momentum, due to the \textit{primordial} transverse momentum of the struck quark inside the nucleon.  
The ``lightcone''  components of $p$ are  $p^{\pm}=p^0\pm p^z$ and the transverse ones $\pt=(p^x,p^y)$ (similarly for $k$ and $k'$). 
The mass-shell constraint writes $p^+p^-=m_h^2+\pt^2\equiv\epsilon^2_h$, where $\epsilon_h$ is the hadron transverse energy.

The energy-momentum sharing between $h$ and $q'$ in Eq. (\ref{eq: splitt elementary}) is drawn at random following the \textit{splitting distribution}  
\be  \label{split-dis} 
dN(q\to h+q') = F_{q\to h+q'}(Z,\textbf{p}_{\rm T},\kv_{\rm T}) \,\frac{dZ}{Z}\,d^2\pt \,,
\ee
where the \textit{longitudinal splitting variable} $Z=p^+/k^+$ is the fraction of forward lightcone momentum of $q$ taken by the hadron $h$. ${(dZ/Z)}\,d^2\pt = d^3\pv/p^0$ is relativistically invariant. 
%

\section{The string + $\mathbf{^3P_0}$ model}

\subsection{Review of the spinless string fragmentation model}

Hadronization of a quark pair $\QA\QbB$ is considered as successive breakings  
of a massive string stretching between $\QA$ and $\QbB$, which we call here a \emph{sting}. 
Each breaking creates a new quark-antiquark pair. A semi-classical treatment of this process leads to a recursive model with a very specific form of the splitting function. We will start with the simple classical $(1+1)D$ \textit{yoyo model} \cite{A-Men} where the created quarks have no mass, no spin and no transverse momentum. Then the complexity will be increased step by step by introducing masses, transverse momenta and spin.

\subsubsection{The (1+1)D yoyo model} 

\begin{figure}[tb]
 \begin{minipage}{.4\textwidth}
  \includegraphics[width=0.9\linewidth]{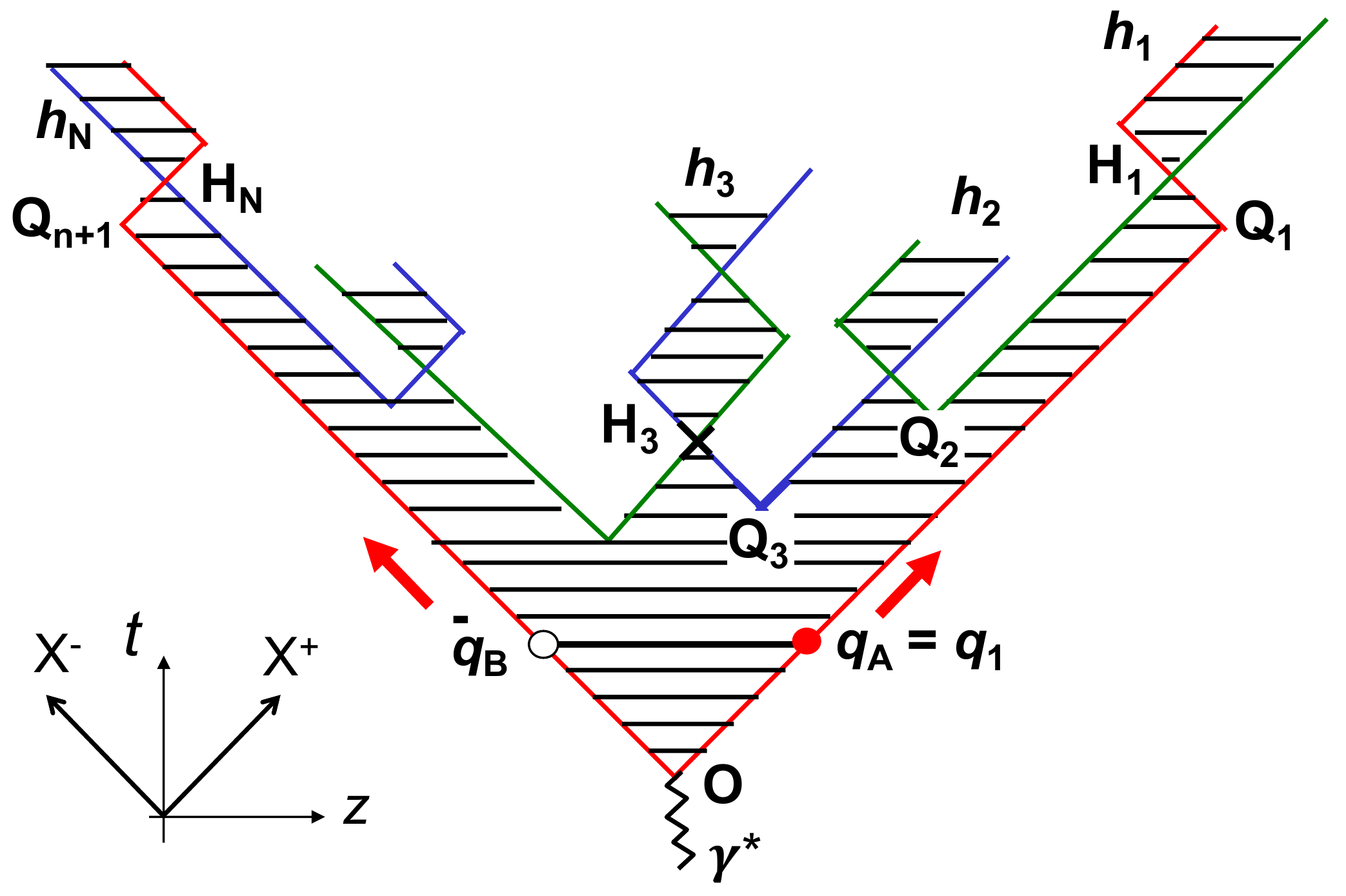}
  \end{minipage} 
 \caption{\small{Space-time history of the string fragmentation. It corresponds to the multiperipheral diagram shown in Fig. \ref{fig:multiperipheral}.}}
  \label{fig:space-time_history}
\end{figure}

In this model everything occurs in the $(t,z)$ hyperplane. One assumes that the \emph{sting} has a uniform probability ${\cal P} \,dz \,dt$ to break in the space-time area $dz\, dt$.  
From the quantum point of view, the ``string fragility'' ${\cal P}$ is taken into account by adding an imaginary part $-i\hbar\Pcal/2$ to the string tension $\kappa\simeq$ 1 GeV/fermi \cite{XA1984,BA-Hof,QYD}.
The complex string tension $\kappa_{\rm C}= \kappa-i\hbar\Pcal/2$ is analogous to the complex mass $m-i\hbar\gamma/2$ of an unstable particle.  
The decay products are small strings which oscillate like yoyos.
Figure \ref{fig:space-time_history} shows the corresponding space-time history.
The string world sheet (hatched domain) is bordered by quark world lines. 
The breaking points $\R_2, \R_3, \cdots \R_N$, completed by the return points $\R_1$ and $\R_{N+1}$, form an a-causal chain, \ie, the 2-D vector $\overline{\R_{r}\R}_{s}$ is space-like. The one-point breaking density in the $(t,z)$ plane is
\be \label{1-dens} 
dN/d^2\R = {\cal P} \, \exp(-{\cal P}\, \overline{{\bf O}\R}^2/2 ) \,. 
\ee
%
where the exponential is the probability that no string breaking occured in the past light cone of $\R$.

Breaking at $\R_{r}$ creates a quark pair $\{q_r\bar q_r\}$. 
~$q_r$ and $\bar q_{r+1}$ meet at point $\Hv_r$ to form the yoyo $h_r$, which represents a stable hadron or a resonance, depending on its mass. Its 2-momentum $p_r$ is given by

\be \label{p}
\check p_{r} = \kappa \, \overline{\R_{r+1}\R}_{r}\,,  
 \ee
where the "check" interchanges the time and $z$ components, \ie,  $\check v = (v^z,v^0) $  for any vector $v=(v^0,v^z)$. In principle, a yoyo $h_r$ can further break, simulating the decay of a resonance. However one limits the model to the production of these ``primary'' hadrons. It already reproduces salient properties of the hadronic final states: two back-to-back jets, Feynman scaling, charge retention effect and rapidity plateau.
A weak point of the model is its fully continuous mass spectrum starting at $m=0$.%

\subparagraph{\it{Recursive treatment.}} 
 
In the ``$P_z=-\infty$'' frame, or using $X^-=t-z$ as time variable, the hadron emission points $\Hv_1, \cdots \Hv_N$ are ordered in time according to their ranks in Eq. (\ref{recursif}). This allows to treat the \textit{sting} decay as a recursive quark fragmentation, identifying $q_r$ of the $q_r\bar q_r$ pair with $q_r$ of Eq. (\ref{recursif}).
 \begin{figure}[tb]
 \begin{minipage}{.4\textwidth}
  \includegraphics[width=0.9\linewidth]{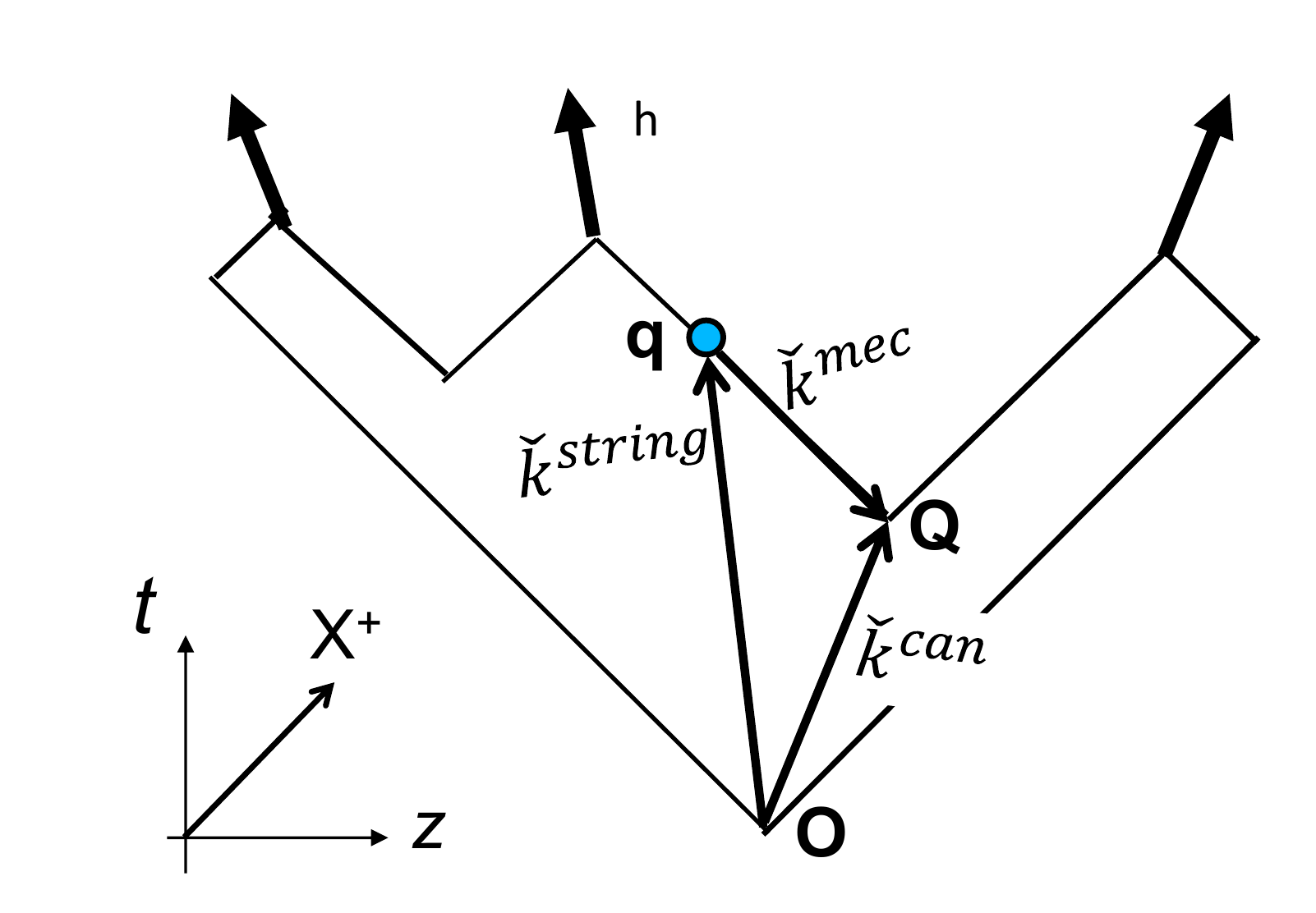}
  \end{minipage}
 \caption{\small{Pictorial representation of the canonical and mechanical momenta. The symbol $\check{}$ represents the symmetry of the vector with respect to the $X^+$ axis.}}
  \label{fig:canonical}
\end{figure}
In the string picture the 4-momentum conservation $k_r=p_r+k_{r+1}$ applies to the \textit{canonical} momentum of the quark given by
\be \label{k}
\check k^{\rm can}_{r} = (k_r^z,k_r^0) = \kappa \, \overline{{\bf O}\R}_r \,.   
 \ee
At any point {$\qv$} of the $r^{\rm th}$ quark trajectory we have
\be \label{k'}
 k^{\rm can}_{r} = k_r^{\rm mec} + k_r^{\rm string} \,,
 \ee
where $k_r^{\rm mec}$, given by $\check k_r^{\rm mec}=\kappa\, \overline{\qv\R}_r $,
 is the \textit{mechanical} momentum of the quark (see Fig. \ref{fig:canonical}). It is on-mass-shell, but varying along the quark line. 
$k_r^{\rm string} $ 
is the momentum flow, from right to left, of the string through {${\bf O}\qv$} and plays the role of a linear 2-potential. It is given by $\check k_r^{\rm string} = \kappa \, \overline{{\bf O} \qv}$.

Any recursive model can be uniquely defined by the \textit{double density of consecutive quarks} in momentum space. In the (1+1)D yoyo model it is
 \be \label{yoyo1} 
 \frac{dN_{q'q}}{d^2k' \, d^2k} =   
 (2 \bl)^2 \, \exp(-\bl |k^+k'^-|)   
 \ee
 {with $\bl=\Pcal/(2\kappa^2)$ and $k\equiv k^{\rm can}$. 
Note that $k^+>0$ and $k^-<0$ for all quarks, except for $k\A^- = k\B^+ = 0$. 
The exponential factor is the probability that no string breaking occurs in the past light cone of $\Hv$, as shown in the right picture of Fig. \ref{fig:vertex}.

 \begin{figure}[tb]
  \begin{minipage}{.4\textwidth}
   \includegraphics[width=0.9\linewidth]{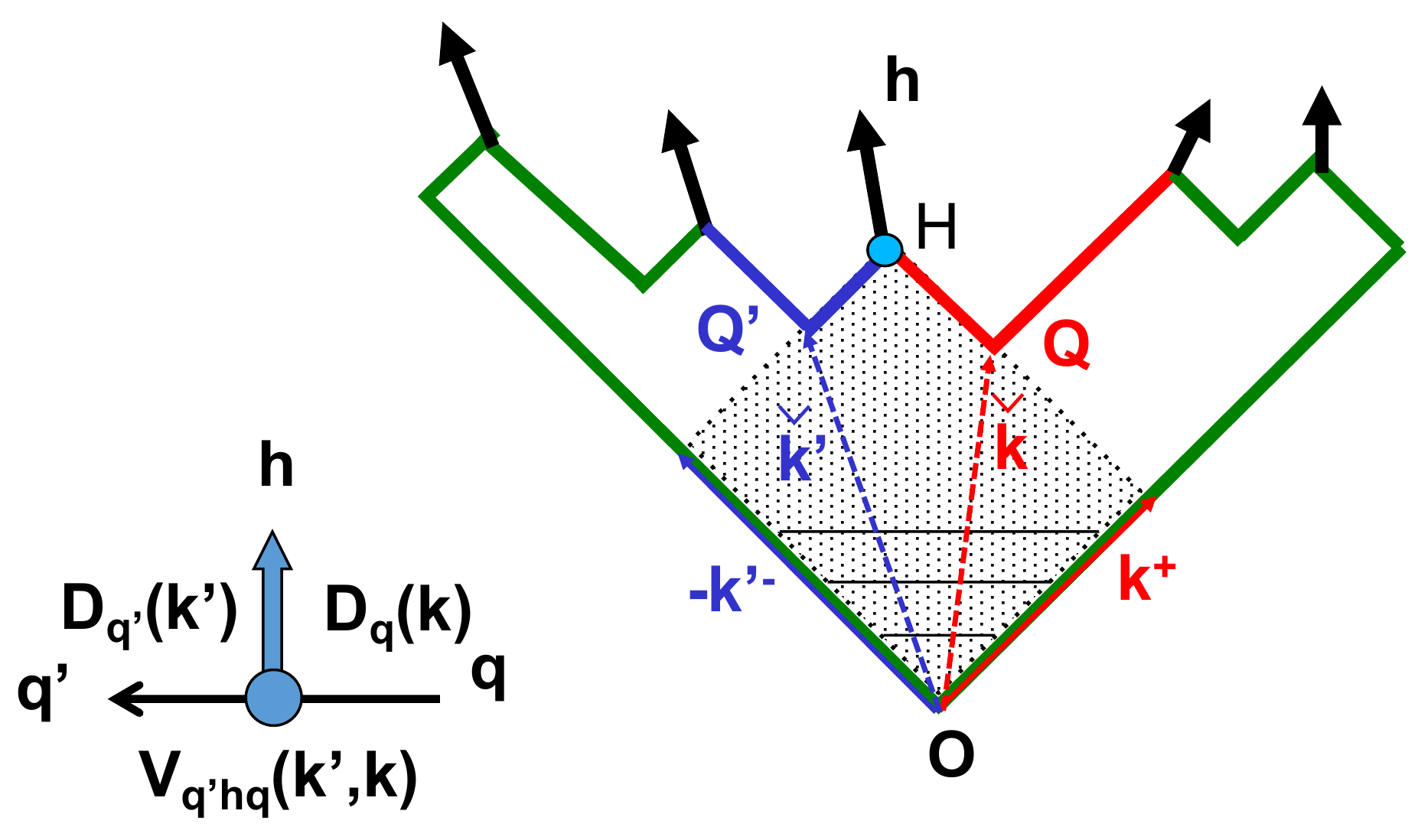}
   \end{minipage} 
 \caption{\small{Representation of the emission vertex of a hadron $h$ in the string fragmentation picture (right) and in the multiperipheral picture (left).}}
  \label{fig:vertex}
\end{figure}

 From Eq. (\ref{yoyo1}) we obtain the single quark density
 \be \label{yoyo2}
 {dN_{q}}/{d^2k} =  2 \bl \, \exp(-\bl |k^+k^-|),
\ee
equivalent to Eq. (\ref{1-dens}), and the splitting function
\begin{eqnarray} \label{yoyo3}
\nonumber \frac{dN_{q\to h+q'} }{d^2k} &=& \bigg(\frac{dN_{q'q}}{d^2k' \, d^2k}\bigg) \bigg(  \frac{dN_q}{d^2k} \bigg)^{-1} \\
&=& 2 \bl \, \exp(-\bl m_h^2/Z) \,.
\end{eqnarray}

\subparagraph{\it{Multiperipheral feature.}} 
The (1+1)D yoyo model can be cast in the form of a multiperipheral model with quark exchanges, as pictured in Fig. \ref{fig:multiperipheral}. In Fig. \ref{fig:vertex} a vertex is represented both in the string and in the multiperipheral picture. The squared vertex function is
\be \label{yoyo1'}
|\vers(k',k)|^2 = {dN_{q'q}}/{(d^2k' \, d^2k}) 
\ee
and the squared propagator
\be \label{yoyo2'}
|\pros(k)|^2 = ({dN_{q}}/{d^2k})^{-1}. 
\ee
Equation (\ref{yoyo2}) insures the cutoff in the quark virtuality $-k^2 = |k^+k^-|$.

A connection between a QCD multiperipheral model and the string fragmentation model is also discussed in Ref. \cite{collins-rogers}. They point out a non-conservation of quark momentum which may be related to that of $k_r^{\rm mec}$ defined in Eq. (\ref{k'}).

\subsubsection{Introducing transverse momenta and actual hadron masses: the Lund symmetric splitting function (LSSF)} 

{Classically, string breaking can only create massless quarks with zero transverse momenta. 
To overcome this limitation one assumes that quarks pairs are created by a tunneling mechanism analogue to the Schwinger mechanism of $e^+e^-$ pair creation in a strong electric field. The quark and the antiquark of the $r^{\rm th}$ pair have then opposite transverse momenta, $\kv_{r{\rm T}}$ and $-\kv_{r{\rm T}}$, which are absorbed by hadrons $h_r$ and $h_{r-1}$. 
Figure \ref{fig:space-time_history} is then} an approximate classical picture, a point $\R_r$ representing only the middle of a tunneling path as shown more in detail in Fig. \ref{fig:tunnel}.
 \begin{figure}[tb]
  \begin{minipage}{.4\textwidth}
    \includegraphics[width=0.9\linewidth]{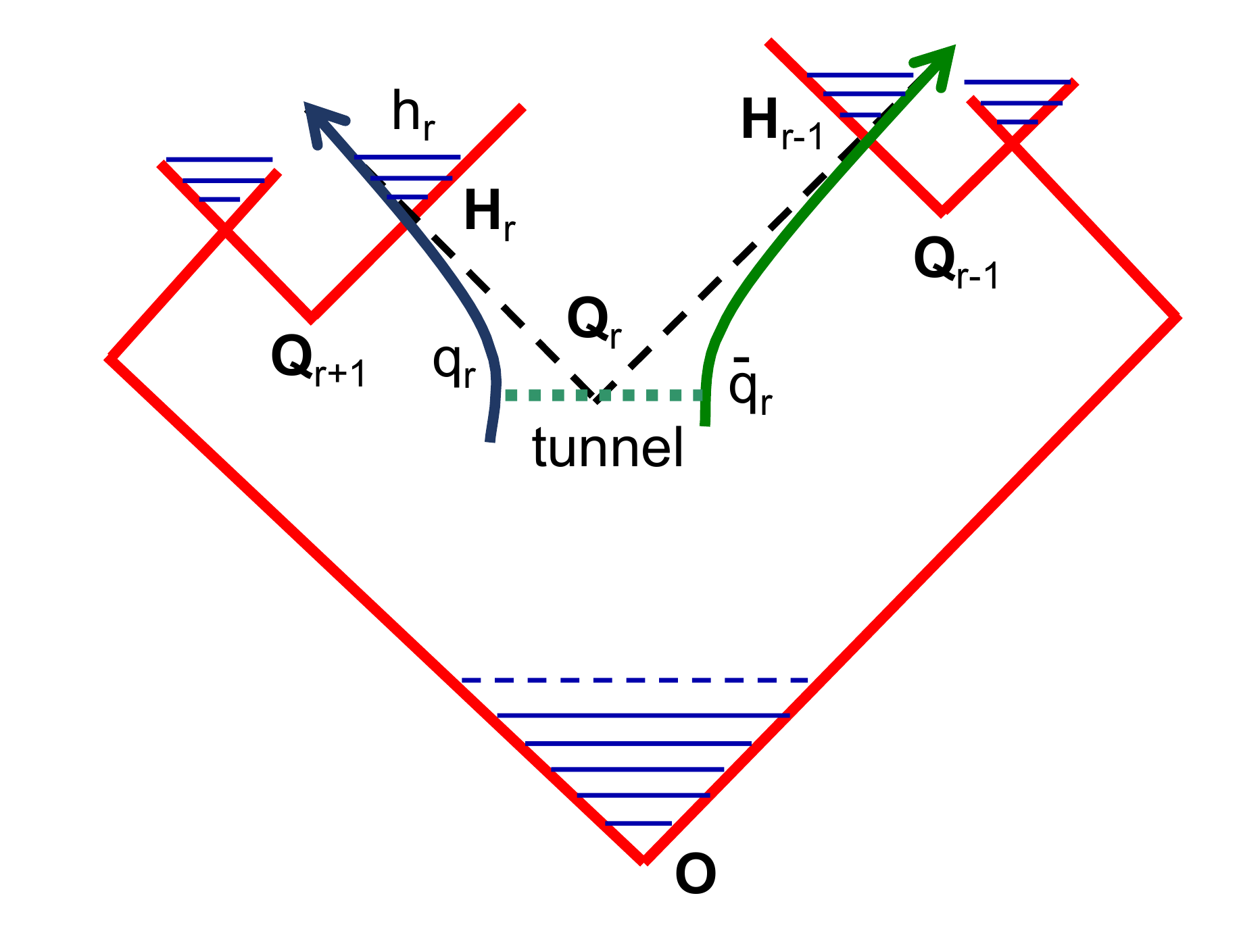}
    \end{minipage}
 \caption{\small{Tunneling process of a $q_r\bar{q}_r$ pair.}}
  \label{fig:tunnel}
\end{figure}
Giving to the hadrons their actual masses and using the principle of ``Left-Right'' symmetry or ``quark Line Reversal'' (hereafter referred to as ``LR symmetry''), the authors of \cite{lund} came to the 
\textit{Lund symmetric} splitting function (LSSF), %
\begin{eqnarray}
\label{LSSF} 
& & F_{q\to h+q'} (k,p) =  \exp( -\bl {\epsilon_h^2}/{Z})  
 \nonumber \\ 
& & \times \left(1/Z-1\right)^{a_{q'}}  
\left({Z}/{\epsilon_h^2} \right)^{a_{q}} 
\nonumber \\ 
 & &\times \ w_{q',h,q} (\kptkpt,{\pt}^2,\kt^2) / u_q(\kt^2) \,.
\end{eqnarray}
The inputs of the model are the parameters ${a_{q}}$ and the function $w_{q',h,q}(\kptkpt,\ptpt,\ktkt)$, which depends on the quark flavors $q$ and $q'$, the hadron species $h$ and the transverse momenta. 
In the most general model $a_{q}\equiv a_{q}(\ktkt)$. 
$u_q(\ktkt)$ is a normalizing factor.
For LR symmetry $w$ must be symmetrical under $\{q,\ktkt\} \rightleftharpoons \{q',\kptkpt\}$ together with $h\rightleftharpoons\bar h$.


Like the (1+1)D yoyo model, the Lund symmetric model can be cast in a multiperipheral form represented in Fig. \ref{fig:multiperipheral}.
Equations (\ref{yoyo1}-\ref{yoyo2'}) for the quark densities, the vertex, the propagators and the splitting function become 
 \be \label{vertex1} 
 \frac{dN_{q'hq}}{d^4k' \, d^4k} =   2\delta(p^2-m_h^2) \,  |\vers_{q'hq}(k',k)|^2 \,,
 \ee
 \begin{eqnarray}
 |\vers_{q'hq}(k',k)|^2 =   \left({k'^+}/{p^+}\right)^{a_{q'}} 
 \exp(-\bl |k'^-k^+|) \,
 \nonumber\\
 \times \left|{k^-}/{p^-}\right|^{a_{q}} 
 w_{q',h,q} (\kptkpt,\ptpt,\ktkt) \,,  
 \label{vertex2} 
 \end{eqnarray}
 \begin{equation}  \label{propag} 
 |\pros_q(k)|^{-2} = {U}_q(k) \equiv {dN_q}/{d^4k} 
 \end{equation}
\begin{equation}\label{Ustring}
U_q(k) = u_q(\ktkt) \, \exp\left(\bl \,|k^+k^-| \right) \, |k^+k^-|^{a_{q}} 
 \end{equation}
 \begin{eqnarray}
 u_q(\ktkt)  &=& 
 \sum_h \int d^2{\kpt} \ w_{q',h,q} (\kptkpt,\ptpt,\ktkt)  \int \frac{dZ}{Z} 
 \nonumber \\
 & \times &  
 \left({Z}/{\epsilon_h^2}\right)^{a_{q}}  (1/Z-1)^{a_{q'}} 
 \,  \exp( -\bl \,\epsilon_h^2/Z) \,,  
 \label{ustring} 
 \end{eqnarray}
 \be \label{split}
 \frac{dN_{q'hq}/d^4k' \, d^4k}{dN_q/d^4k} = 2\delta(p^2-m_h^2) \, F_{q\to h+q'} (k,p) \,,
 \ee
where $F$ is given by Eq. (\ref{LSSF}). $w_{q',h,q}$ is normalized such that any time-like curve passing by ${\bf O}$ in Fig. \ref{fig:space-time_history} is crossed by one and only one $\R\R'$ segment. 
The power-law factors lead to a multi-Regge behavior for large rapidity gaps \cite{XA1984}. 
In a semiclassical approach, the quantum actions of the quarks produce such factors, 
with $a_{q}(\ktkt) = \alpha_{\rm out} - \bl \,(m_q^2+\ktkt)$ \cite{DS11}. 
Note that the vertex and the propagator are not invariant under the full Lorentz group, but under the subgroup generated by 

\ni ~(a) the rotations about $\zu$,

\ni ~(b) the Lorentz boosts along $\zu$,

\ni ~(c) the reflection about any plane containing $\zu$.  

\ni Indeed, the string axis defines a privileged direction of space.

We take $w$ of the form
\begin{eqnarray}
w 
= |C_{\rm q',h,q} \ \check g(\epsilon_h^2) \,
f\T(\kptkpt) \,  f\T(\ktkt) |^2 \,.
\label{wstring} 
\end{eqnarray}
$C_{q',h,q}$ is proportional to the $(\bar q' q)$ wave function in flavor space. 
It acts upon the hadron species distribution. 
$f\T(\ktkt)$ is a fast decreasing function of $\ktkt$  (\textit{e.g.}, a Gaussian). 
$\check g(\epsilon_h^2)$ acts upon the correlation%
\footnote{Such correlations are present in the standard multiperipheral model, where $\langle\kpt\rangle = (1-Z) \langle\kt\rangle$ \cite{LUG}.}
between $\kt$ and $\kpt$, since $\epsilon_h^2=m_h^2+(\kt-\kpt)^2$.  
For $\check g(\epsilon_h^2)=1$ one obtains $\langle\kt\cdot\kpt\rangle>0$ due to the factor $\exp( -\bl {\epsilon_h^2}/{Z})$ in Eq. (\ref{LSSF}). 
In PYTHIA, such a correlation is absent, due to the particular choice 
$\check g^2(\epsilon_h^2) = 1/ N_a(\epsilon_h^2)$ 
with
\be \label{Na} 
N_a(\epsilon_h^2) = \int_0^1 \frac{dZ}{Z} \, \left(\frac{1-Z}{\epsilon_h^2}\right)^a \, \exp( -\bl \,\epsilon_h^2/Z)
\ee
where $a_q(\ktkt)=a$ is taken flavour- and $\ktkt$- independent.
%

\subsection{The classical \textit{string} + $\boldsymbol{^3P_0}$ mechanism}      
One assumes that the string breaking, in which a $q_r\bar q_r$ pair is created ($r$ is the rank of the splitting), occurs via a tunnel effect and that, at the end of this process, $q_r$ and $\bar q_r$ are on the string axis (the $z$ axis), with transverse momenta $\kv_{r{\rm T}}$ and $-\kv_{r{\rm T}}$ respectively, zero longitudinal momenta, and separated by the vector 
\be \label{dqqbar}
{\bf d}_r \equiv \rv_{q_r}-\rv_{\bar q_r}=-2\,\zu\,(m_{q_r}^2+\kv_{r{\rm T}}^2)^{1/2}/\kappa  \,.
\ee
The string between $q_r$ and $\bar q_r$ has been ``eaten'' by the pair. 
The modulus of ${\bf d}_r$ is fixed by energy conservation and its orientation is that of the initial color flux, \ie, from $q\A$ to $\bar q\B$. 
The quark pair has a relative orbital momentum 
\be \label{Lqqbar}
\Lv_r={\bf d}_r\times\kv_{r{\rm T}} \,.
\ee
One furthermore assumes that the $q_r\bar q_r$ pair is in the $^3P_0$ state (which possesses the quantum numbers of the vacuum).  In such a state the spins are parallel and opposite to $\Lv_r$~:
\be \label{correl3P0} 
\langle \sv_{q_r} \cdot\sv_{\bar q_r} \rangle>0
\,,\quad  
\langle \sv_{q_r} \cdot\Lv_r \rangle <0
\,,\quad  
\langle \sv_{\bar q_r} \cdot\Lv_r \rangle <0 \,. 
\ee
It follows from (\ref{dqqbar}), (\ref{Lqqbar}) and (\ref{correl3P0}) that the polarisations of $q_r$ and $\bar q_r$ are correlated to their transverse momenta:
\be \label{correl-ks} 
\langle\kv_{r{\rm T}}\times\sv_{q_r}\rangle\cdot\zu>0 \,,\quad \langle\kv_{r{\rm T}}\times\sv_{\bar q_r}\rangle\cdot\zu>0 \,.
\ee
Besides (\ref{correl3P0}), which correlates $\sv_{q_r}$ and $\sv_{\bar q_r}$, there is a correlations between $\sv_{q_r}$ and $\sv_{\bar q_{r+1}}$ coming from the internal wave function of the meson $h_r$. In particular, if $h_r$ is a pseudoscalar meson ($\pi,\,K,\,\eta$ or $\eta'$),
\be \label{correl-hr}
\langle\sv_{q_r}\cdot \sv_{\bar q_{r+1}}\rangle<0 \,, 
\ee
as required by the $^1S_0$ internal wave function.

\begin{figure}[tb]\centering
\begin{minipage}{.5\textwidth}
  \includegraphics[width=1.0\textwidth]{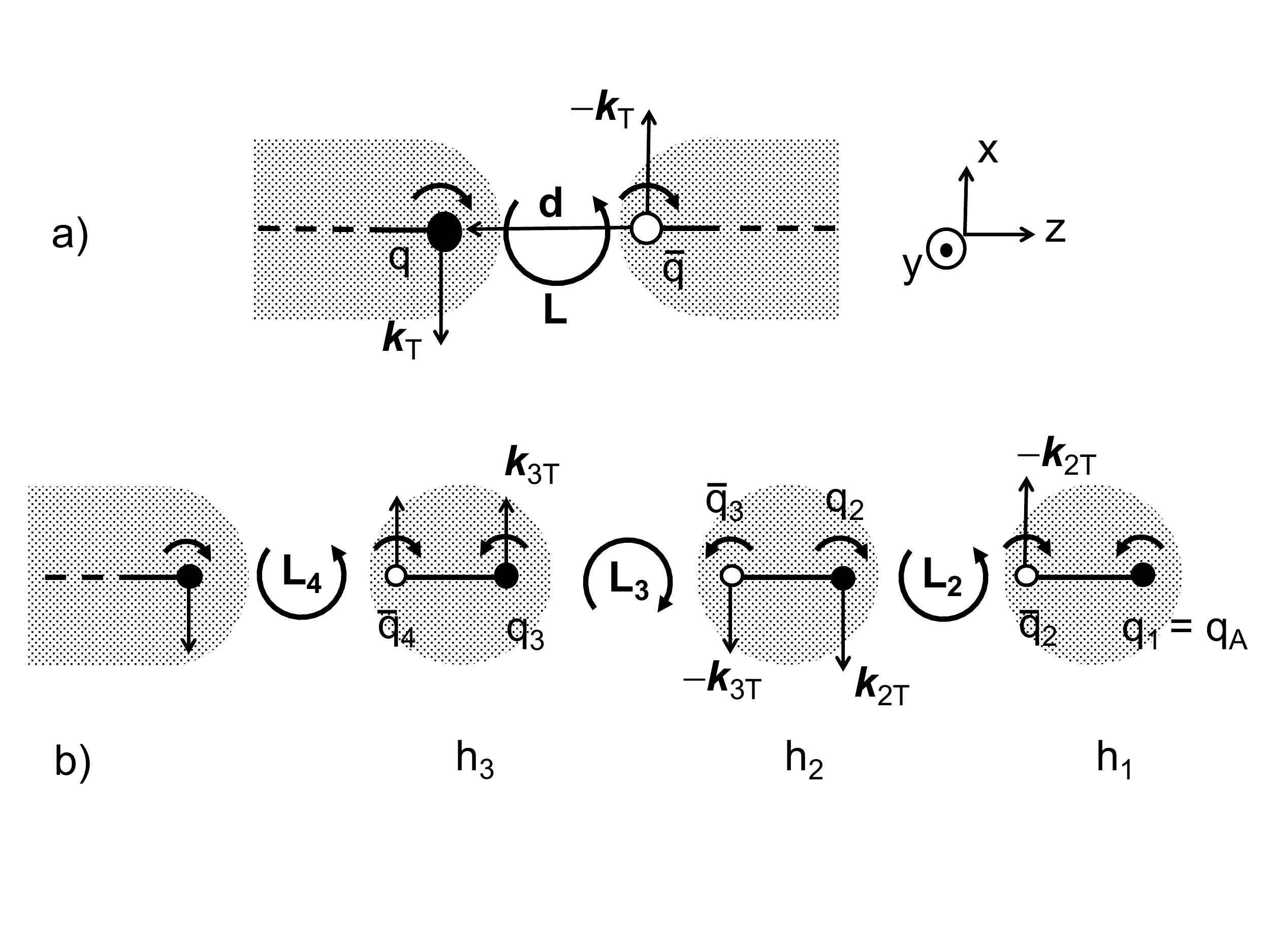}
\end{minipage} 
\caption{\small{Classical $string+ {}^3P_0$ mechanism of Collins effect. (a) Elementary mechanism.
(b) Iteration in the emission of pseudo-scalar mesons.}}\label{fig:3P0}
\end{figure}

Figure \ref{fig:3P0} depicts the spin and $\kt$ correlations in the recursive decay of the \textit{sting} when only pseudoscalar mesons are emitted and assuming that $q_A$ is polarized along $+\yu$ (as represented by an anti-clockwise arrow).
According to (\ref{correl3P0}) and  (\ref{correl-hr}), $q_2$ and $\bar q_2$ are both polarized along $-\yu$ (clockwise arrow) and their relative orbital momentum ${\bf L}_2$ is along $+\yu$  (anti-clockwise arrow).
Then
$\bar q_2$  and  $q_2$  move respectively in the $+\xu$ and $-\xu$ directions, in accordance with (\ref{correl-ks}).
The transverse momentum $-\kv_{2{\rm T}}$ of $\bar q_2$, which is toward $+\xu$, is absorbed by $h_1$, 
resulting in a Collins effect with $\langle p_{1,x}\rangle>0$, more generally
  $\langle\pv_{1{\rm T}}\times\Sv_{{\rm A,T}}\rangle\cdot\zu>0$.

\subsection{Quantum treatment of the quark spin}      

We encode the quark spin degree of freedom with Pauli spinors and, using the multiperipheral approach, transform the vertex function $\vers$ and the propagator $\pros$ of Eqs. (\ref{vertex2})-(\ref{propag}) into {2$\times$2} matrices acting on quark spin. 
$w$, $u$ and the quark density $U$ of Eqs. (\ref{vertex2})-(\ref{ustring}) become density matrices (Hermitian and semi-positive definite) in spin space.  
Full Lorentz invariance would require the use of Dirac spinors, but Pauli spinors are sufficient to satisfy the invariance under the above mentioned subgroup.
Note that it does not take into account the whole spin information (2 q-bits) carried by an off-mass-shell Dirac particle. 

\subsubsection{General formalism}   

We first consider a general mutiperipheral model, not necessarily combined with the string model.
The amplitude for reaction (\ref{hadronization}) is
\begin{eqnarray}\label{multiperampli} 
& & \langle \Sv\B | \Mcal(\q\A\bar\q\B \to \h_1\h_2...\h_N)| \Sv\A \rangle 
= \nonumber \\
& &\langle \Sv\B  |\, \pros(\q\B) \, \vers(\q\B,\h_N,\q_{N})\, \pros(\q_{N}) 
\cdots  \nonumber \\  
& \cdots&\vers(\q_3,\h_2,\q_{2}) \, \pros(\q_2) \,\vers(\q_2,\h_1,\q\A) \, \pros(\q\A) \, | \Sv\A  \rangle.
\end{eqnarray}
To save place, the gothic letters gather several variables: for a quark $\q=\{q,k\}$, where $q$ is the flavor; for a hadron $\h=\{h,p,s_h\}$, where $h$ is the hadron species and $|s_h\rangle$ belongs to an adopted spin basis (\textit{e.g.}, helicity basis). Thus, $\pros(\q)\equiv\pros(q,k)$ and $\vers(\q',\h,\q)\equiv \vers_{q',h,s_h,q}(k',k)$.  
$| \Sv \rangle$ is the Pauli spinor of polarization $\Sv=({\Sv_{{\rm T}},S_{\rm L}})$, with T and L referring to the transverse and longitudinal polarizations of the quark respectively. $| \Sv\B  \rangle$ is related to the polarization $\Sv_{\overline{\rm B}}$ of the antiquark $\bar q\B$ by 
\be
| \Sv\B  \rangle = - \sigma_z | - \Sv_{\overline{\rm B}}   \rangle,
\ee
which is the analog of the Dirac spinor $v(k,\Sv)  = - \gamma_5 \, u(k,-\Sv)$ of an antiparticle. 
%

The functions $\pros(\q)$ and $\vers(\q',\h,\q)$ may be chosen as input of the model. However 
they can be ``renormalized'' by the transformation 
\begin{eqnarray}
\pros(\q) &\to& \Lambda_L(\q) \, \pros(\q) \, \Lambda_R(\q) \,,
\nonumber \\
\vers(\q',\h,\q) &\to& \Lambda_R^{-1}(\q') \, \vers(\q',\h,\q) \, \Lambda_L^{-1}(\q)
 \label{renorm}
\end{eqnarray}
without changing the physical result, so different inputs lead to the same model.
Here we use the ``renormalized input'' method of \cite{DS13}, where the $4\times 4$ matrix
\begin{eqnarray}\label{eq: new N}
\nonumber \langle j\,j'|{\cal N}(\q',\h,\q)|i\,i'\rangle = 2\delta(p^2-m^2)\, \langle j\,j'|\vers^\dagger_{\rm pt}|\h\rangle \ \langle \h|\vers_{\rm pt}| i\,i'\rangle\\
\end{eqnarray}
is the \textit{density operator} of pairs of consecutive quarks. $i,j$ label spin the states for $q$ and $i',j'$ label the spin states for $q'$. In Eq. (\ref{eq: new N}) we have introduced the partially transposed matrix
\begin{equation}
\langle \h|\vers_{\rm pt} |i\,i'\rangle \equiv \langle i' |\vers(\q',\h,\q)|i\rangle.
\end{equation}
Equation (\ref{eq: new N}) generalizes Eq. (\ref{vertex1}). 
In fact ${\cal N}$ is a \textit{density matrix} in spin space but a classical density in momentum space, since we still treat momenta classically (see footnote 1).  
The single-quark density operator, generalizing Eq. (\ref{propag}), is
\be \label{U} 
\frac{1}{\pros (\q)  \, \pros^\dagger(\q)}  
=  U(\q) = \sum_{h,s_h} \int\frac{d^3\pv}{p^0} 
\,\vers^\dagger(\q',\h,\q)\ \vers(\q',\h,\q) \,.
\ee
Invariance under reflection about the $(x,z)$ or $(y,z)$ plane requires
\begin{eqnarray}
U(\q) = U_0(\q) + U_1(\q) \, \vecsigma\cdot\ntil(\kv) \,, 
\nonumber \\
\pros(\q) = \pros_0(\q) + \pros_1(\q) \, \vecsigma\cdot\ntil(\kv) \,. 
\label{reflex}
\end{eqnarray}
where $\ntil(\kv) \equiv \zu\times\kt/|\kt|$. 
$U$ is Hermitian and semi-positive definite: $U_0(\q)$ and $U_1(\q)$ are real with $U_0(\q) \ge  |U_1(\q)|$. 
We assume the strict inequality so that a solution of Eq. (\ref{U}) for $\pros (\q) $ exists. 
$\pros (\q) $ is not uniquely determined by Eqs. (\ref{U}) and (\ref{reflex}) but, using the ``renormalization'' (\ref{renorm}), we can take the positive definite solution $\pros (\q) = U^{-1/2}$ without loss of generality. 

Introducing the \textit{splitting matrix}  
\be   \label{Tmat}
T(\q',\h,\q) \equiv \vers(\q',\h,\q) \, \pros(\q) \,,
\ee
the \textit{polarized} splitting function to be used in Eq. (\ref{split-dis})   becomes
\begin{eqnarray}  
\label{polar-split-mat}   
F_{q',h,q}(Z,\kpt,\kt)
=  \Tr \left[  T(\q',\h,\q) \ \rho(q)  \ T^\dagger(\q',\h,\q) \right] \,. ~
\end{eqnarray}
where $\rho(q) = (\un + \Sv_q \cdot \vecsigma)/2$ is the spin density matrix of quark $q$, normalized to unit trace. 
$F$ obeys the normalization condition 
\be \label{tir} 
 \sum_{h, s_h} \, \int \frac{d^3\pv}{p^0} \, 
F_{q',h,q}(Z,\pt,\kv_{\rm T}) = 1 \,.
\ee
%

%
%
From the practical point of view Eq. (\ref{polar-split-mat}) is used to draw the species $h$, the spin state $s_h$ and the momentum $\pv$ of hadron $h$.

The spin density matrix of the left-over quark $q'$ is 
	\begin{equation}\label{repol} 
	\rho(q') = 
	\left[T(\q',\h,\q) \ \rho(q) \ T^\dagger(\q',\h,\q) \right] / \Tr \, [\text{idem}] \,.
	\end{equation}}
Thus Eqs. (\ref{polar-split-mat}) and (\ref{repol}) are the basis for the recursive generation of a polarized quark jet.
	
	%



\subsubsection{Combination with the string model}

For the vertex $\vers$ we take
\begin{eqnarray}
\vers(\q',\h,\q) &=&  \left({k'^+}/{p^+}\right)^{a_{q'}/2} 
\, \exp(-\bl |k'^-k^+|/2)   
\nonumber \\
&\times&
|{k^-}/{p^-}|^{a_{q}/2} \, g(\q',\h,\q) \,,
\label{SMvertex}
\end{eqnarray}
which generalizes Eq. (\ref{vertex2}). $g(\q',\h,\q) = g_{q',h,s_h,q}(\kpt,\qt)$ is a 2$\times$2 matrix acting on quark spin. It also contains the flavor and $\kt$ dependence of $\vers$.%
\footnote{We omitted a spin-independent phase factor coming from the string action. Indeed, the recursive model is based on the ladder approximation of the multiperipheral model. In this approximation the spin independent phases of the amplitudes are irrelevant.}

We decompose Eq. (\ref{U}) in 
\be
\pros(\q) = |k^-k^+|^{a_{q}/2}  \, \exp(\bl \,|k^-k^+|/2) \,  d_q(\kt) \,,
\ee
\be \label{dud} 
\left[d_q(\kt) \, d^\dag_q(\kt) \right]^{-1} = u_q(\kt)  \,, 
\ee  
\begin{eqnarray}
u_q(\kt)  =
\sum_h\sum_{s_h} \int d^2{\kpt}  \ g^\dagger(\q',\h,\q)\,g(\q',\h,\q)  
\nonumber \\
\times \int_0^1 \frac{dZ}{Z} \, \left(\frac{1-Z}{Z}\right)^{a_{q'}} 
\left(\frac{Z}{\epsilon_h^2}\right)^{a_{q}} 
\, \exp( -\bl \,\epsilon_h^2/Z)\,.
\label{u'} 
\end{eqnarray}
$w$ of Eq. (\ref{wstring}) has been replaced by the rank-1, semi-positive matrix  $g^\dagger\,g$.  
%

\paragraph{Particular choices of $a_q$ and $g(\q',\h,\q) $.} We take $a_q=a_{q'}=a$ = constant and $g(\q',\h,\q) $ of the form	
\begin{eqnarray}
g(\q',\h,\q) &=& C_{\rm q',h,q} \ \check g(\epsilon^2_h) \,
\nonumber \\  &\times& 
\g_{q'}(\kpt)  \,   \Gamma_{h,s_h}(\kpt,\kt)   \, \g_q(\kt)  \,, 
\label{factorig} 
\end{eqnarray}
with 
\be   \label{gamma(q)}
\g_q(\kt)   =  (\mu_q + \sigma_z\, \vecsigma\cdot\kt) \, f\T(\ktkt) \,. 
\ee
The term $\mu_q + \sigma_z\, \vecsigma\cdot\qt$ is the 2$\times$2 analogue of the numerator $m_q + \boldsymbol{\gamma}\cdot k$ of the Feynman propagator. The function $f\T$ provides the cutoff in $k\T$. Inspired by the Schwinger mechanism, we take the gaussian $f_{\rm T}(\ktkt)=\exp(-\bt\ktkt/2)$ with $\bt$ a free parameter. $\mu_q$ is a complex parameter having the dimension of a mass. With $\IM(\mu_q)>0$, the factor $\mu_q + \sigma_z\, \vecsigma\cdot\qt$, introduced in \cite{DS09}, reproduces the classical string+$^3P_0$ mechanism.
$\Gamma$ is a 2$\times$2 matrix which depends on $\kt$ and $\kpt$ at most as a polynomial. We restrict ourselves to pseudoscalar mesons and, to zero order in $\kt$ and $\kpt$, we take
%
\begin{eqnarray} \label{Gamma1}
\Gamma_h &=& \sigma_z\, 
 \label{Gamma3}
\end{eqnarray}
which is the analogue of $\gamma_5$.
In Ref. \cite{spin16} the slightly different choice $\Gamma_{h,s_h}(\kpt,\kt)=\mu\sigma_z+\vecsigma\cdot\pt$ and $\g = \exp(-\bt\ktkt/2)$ was made. It gives practically the same result. Vector and axial mesons can also be introduced as shown in Ref. \cite{DS09}.

%
Using Eq. (\ref{factorig}) we can rewrite Eq. (\ref{u'}) as
%
\be \label{factoru} 
u_q(\kt) =  \g^\dagger_q(\kt)  \, \hat u_q(\kt)  \, \g_q(\kt) \,,
\ee
\begin{eqnarray} 
\hat u_q(\kt) &=&
\sum_{h} |C_{q',h,q}|^2   \int d^2{\kpt} \, \check g^2(\epsilon_h^2)  
\, N_a(\epsilon_h^2) 
\nonumber \\
&\times & 
\sum_{s_h}  \Gamma^\dagger_{h,s_h}  \
\g^\dagger_{q'}(\kpt)\,\g_{q'}(\kpt) \ 
\Gamma_{h,s_h}  
\label{uchapeau} 
\\
&\equiv& \hat u_0(\ktkt) + \hat u_1(\ktkt) \,  \vecsigma\cdot\ntil(\kv) \,, 
\label{integra'''} 
\end{eqnarray}
with $\hat u_0 > |\hat u_1|$. 
$N_a(\epsilon_h^2)$ is given by Eq. (\ref{Na}). As solution of Eq. (\ref{dud}), we take
\be
d_q(\kt) =  \g^{-1}_q(\kt)  \, \hat u^{-1/2}_q(\kt)   \,.
\ee
The splitting matrix in Eq. (\ref{Tmat}) takes the explicit form
\begin{eqnarray}
T &=& 
C_{q',h,q} \  \check{g}(\epsilon_h^2)  \, \Delta_{q'}(\kpt) \,  \Gamma_{h,s_h}  \hat u_q^{-1/2}(\ktkt)
\nonumber\\
&\times& [(1-Z)/\epsilon_h^2]^{a/2} \, \exp[ -\bl \,\epsilon_h^2/(2Z)] \,,
 \label{Tematrix}
\end{eqnarray}
to be used in the algorithm described after Eq. (\ref{tir}). $\hat u_q(\ktkt)$ has to be calculated beforehand.  

\paragraph{Particular choices of $\check g(\epsilon_h^2)$.} 

As in the spinless case, this function acts upon the  \textit{spin-independent} $(\kt,\kpt)$ correlation
which adds to the one mediated by the quark spin.
%
Let us give four examples:

\ni ~a) $\check g(\epsilon_h^2)=(\epsilon_h^2)^{a/2} $,

\ni ~b) $\check g(\epsilon_h^2)=(\epsilon_h^2)^{a/2}\, e^{c\bl \epsilon_h^2}$, with $c\le 1$,

\ni ~c) $\check g(\epsilon_h^2)=1/\sqrt{N_a(\epsilon_h^2)}$, 

\ni ~d) $\check g(\epsilon_h^2)=e^{c\bl \epsilon_h^2}/\sqrt{N_a(\epsilon_h^2)}$, with  $c \le 0$.

\ni Choice a) favors $\kt\cdot\kpt>0$.  
In choice b) this correlation is reinforced for $c<0$ and weakened for $c>0$.
Choice c) suppresses it, like in PYTHIA or in the simplified model of \cite{DS09}. In choice d) the factor $e^{c\bl \epsilon_h^2}$ restores it.

%
%

%


\subsection{The polarized $q\rightarrow h +q'$ splitting function according to the $^{3}P_0$ mechanism}\label{sec:code 2}

We take $\check g(\epsilon_h^2)= (\epsilon_h^2)^{a/2}$, $f_{\rm T}(\ktkt) = e^{-\bt\ktkt/2}$ and a unique complex mass parameter for all quark flavours, \ie, $\mu_q\equiv \mu=(\RE\mu,\IM\mu)$. 
With our choice of $\check g$ two successive quark transverse momenta $\kt$ and $\kpt$ are correlated. 
Gathering Eq. (\ref{polar-split-mat}), Eq. (\ref{gamma(q)}) and  Eq. (\ref{Tematrix})
we obtain the \textit{polarized splitting function} to be used in (\ref{split-dis})
\begin{eqnarray}\label{eq: splitt spin}
\nonumber &\,& F_{q',h,q}(Z,\pt,\kt) = |C_{q',h,q}|^2\,\left[(1-Z)/\epsilon_h^2 \right]^a\\
 \nonumber &\times& \exp(-\bl\epsilon_h^2/Z-\bt\kptkpt)\\
&\times&
  \Tr\left[ (\mu+\sigma_z \vecsigma\cdot\kpt) \Gamma_h \,\hat{\rho}_{int}(q) \,\Gamma^{\dagger}_h(\mu^*+\vecsigma\cdot\kpt\sigma_z) \right] \,~~
\end{eqnarray}
with
\begin{eqnarray} \label{eq:rhoint}
\hat\rho_{\rm int}(q) = \hat{u}^{-1/2}_q(\kt) \, \rho(q) \, \hat{u}^{-1/2}_q(\kt) \,.
\end{eqnarray}
%
The free parameters $a$ and $\bl$ play the same role as $a$ and $b$ in the PYTHIA event generator. They govern the suppressions of $F$ at large and small $Z$ respectively.  
The parameter $\bt$ is linked to the spread of the quark transverse momenta produced at the string cutting points. $\hat{\rho}_{\rm int}(q)$ is an intermediate density matrix which we have not normalized. The corresponding polarization vector is 
\be\label{eq:Sint}
\Sv_{\rm int} = \Tr [ \hat\rho_{\rm int} \, \vecsigma ] / \Tr \hat\rho_{\rm int} \,.
\ee

Working out the trace operations in Eq. (\ref{eq: splitt spin}) with $\Gamma_h=\sigma_z$, the splitting function is explicitly given by
\begin{eqnarray}
\label{eq:splitting explicit}
\nonumber && F_{q',h,q}(Z,\pt,\kt) \propto (1-Z)^a \exp\left(-\bl m_h^2/Z\right)\\
\nonumber &\times& \exp\left( -\bt\xi(Z)\ktkt\right)\exp{\left[-\frac{\bl}{Z\xi(Z)}\left(\kpt-\xi(Z)\kt\right)^2\right]}\\
&\times&\left[|\mu|^2+\kptkpt-2\IM(\mu)\textbf{S}_{int}\cdot\tilde{\textbf{k}'}_{\rm T}\right]
\end{eqnarray}
where $\kpt=\kt-\pt$ and $\xi(Z)\equiv \bl/(\bl+Z \bt)$. The tilde denotes the "dual" of a transverse vector, for instance $\tilde{\textbf{p}}_{\rm T}=\hat{\textbf{z}}\times\textbf{p}_{\rm T}$. The vector $\textbf{S}_{int}$ is the polarization vector of the intermediate spin matrix $\hat{\rho}_{int}(q)$ given in Eq. (\ref{eq:Sint}).



Finally, using Eq. (\ref{repol}) and Eq. (\ref{Tematrix}), the spin density matrix $\rho(q')$ of the quark $q'$ is calculated as
\begin{equation}\label{eq: rho q'}
\rho(q')=\frac{(\mu + \sigma_z \vecsigma\cdot\kt) \Gamma_{h,s_h}\hat{\rho}_{int}(q) \Gamma^{\dagger}_{h,s_h}(\mu^* - \sigma_z\vecsigma\cdot\kt)}{Tr(\,idem\,)}.
\end{equation}

\section{Monte Carlo implementation}\label{sec:code 1}
In this section we describe the simulation code handling the fragmentation of a polarized quark into pseudoscalar mesons ($\pi$, $K$, $\eta^0$ and $\eta'$). It is a stand alone program not yet interfaced with existing event generators. Presently, the flavour and the spin density matrix $\rho(q\A)$ of the fragmenting quark $q\A$ are chosen at the begining of the simulation. The initial quark energy can either be fixed or chosen event by event reading the values from an external file. The output consists in a file with the relevant information on all the hadrons generated in the fragmentation, later on analysed to obtain the azimuthal angle distributions and the analysing powers.

\subsection{The Monte Carlo program structure}
A preliminary task, before starting the generation of the events, is to calculate $\hat u_q(\ktkt)$ from Eq. (\ref{uchapeau}), then $\hat u_q^{-1/2}(\ktkt)$ and tabulate its values. 


The initial kinematics for lepton-proton DIS is defined event by event according to the hard subprocess $l+q_0\to l' + q\A$. We consider the center of mass frame of the system composed by the virtual photon and the target proton. We orient the $\textbf{z}$ axis along the virtual photon momentum and consider first the case where $q\A$ has no primordial transverse momentum $\kt_{\rm prim}$.
This reference frame coincides also with the center of mass frame of the final hadronic system whose lightcone momenta $P^{\pm}$ are defined by
\begin{equation}
P^+=P^-=W
\end{equation}
where $W^2=P^+P^-=(1/x_B-1)Q^2+M^2$ is the squared energy available for the fragmentation process. $M$ is the target proton mass. Thus the reservoir of forward and backward lightcone momenta is fixed for each event by the values of $x_B$ and $Q^2$ taken from samples of real events.

In $e^+e^-$ annihilation $W$ is the center of mass energy and it is the same for all the events of a simulation.

 

In our reference frame the quark $q_A$ travels along the forward lightcone and one can identify $k_A^+\equiv P^+$ (implying $k_{\overline{B}}^+=0$). We only consider the fragmentation of this initial quark and neglect the jet initiated by $\bar{q}\B$, which in the DIS case is a diquark and travels along the backward lightcone with momentum $k_{\overline{B}}^-\equiv P^-$ (implying $k_A^-=0$). 

We simulate the splitting process $q\to h+q'$ recursively, starting with $q= q\A$, following the steps:
\begin{enumerate}
\item generate a new $q'\bar{q}'$ pair
\item form $h=q\bar{q}'$ and identify the type ($\pi, K,\eta^0$ or $\eta'$) of the pseudoscalar meson
\item generate $Z$ according to the $\pt$ integrated splitting function and calculate $p^+=Zk^+$
\item generate $\pt$ according to the splitting function at the generated $Z$ and calculate $\kpt=\kt-\pt$ (for $q_A$, $\kt=\textbf{0}$)
\item calculate $p^-$ imposing the mass shell condition $p^+p^-=m_h^2+\pt^2$
\item test the  exit condition: if it is not satisfied continue to step $7$, otherwise the current hadron is removed and the decay chain ends
\item calculate the hadron four-momentum and store it
\item calculate the spin density matrix of quark $q'$ using Eq. (\ref{eq: rho q'}) with $\Gamma_h=\sigma_z$ and come back to step $1$.
\end{enumerate}

We iterate steps $1-8$ until the exit condition, described below, is satisfied. More details on the different steps are given in the following.

\smallskip
\ni{\it Quark flavor and hadron type generation (steps 1 and 2).}
In step $1$ the generation of $s$ quarks is suppressed with respect to $u$ or $d$ quarks, by taking $P(u\bar{u}):P(d\bar{d}):P(s\bar{s})$ with probabilities $3/7:3/7:1/7$ such that $P(s\bar{s})/P(u\bar{u})=1/3$.

The meson identification at step $2$ uses the isospin wave function and also suppresses the $\eta^0$ meson production with repsect to $\pi^0$ to account for their mass difference. We have choosen $N(\eta^0)/N(\pi^0)\simeq 0.57$ as suggested in Ref. \cite{feynman-fieldw}.

\smallskip
\ni{\it Exit condition (step 6).}
After the $r^{th}$ splitting, the $4$-momentum of the remaining string is $P_{{\rm rem},r+1}=k_{\overline{B}}+k_{r+1}$. Then
\begin{eqnarray}
P^{+}_{rem (r+1)}=P^+_{rem (r)}-p^+_r \\ P^-_{rem (r+1)}=P^-_{rem (r)}-\epsilon_{h_r}^2/p^+_r \\ \textbf{P}_{\rm T,r+1}=\textbf{P}_{\rm T,r}-\textbf{p}_{\rm T,r+1}.
\end{eqnarray}
The remaining squared energy to be used in the generation of the next hadrons is $W^2_{r+1}=P_{rem(r+1)}^+P_{rem (r+1)}^--\textbf{P}_{\rm T,r+1}^2$.
If $P_{rem(r+1)}^-$ becomes negative, the last hadron is rejeceted and a new one is tried. This could happen if the last hadron is generated with a very small value of $Z$.
If $W^2_{r+1}$ falls below a given mass $M_R^2$ the chain terminates and the last hadron generated is erased (exit condition at step $6$). We take $M_R=1.5\, GeV/c^2$ in order to leave enough energy for the production of one baryon, which is not simulated. The observables investigated here are not sensitive to this value.

\smallskip
\ni{\it Recursive splitting (steps $3$ and $4$).}
The energy-momentum sharing in the splitting $q(\kt,k^+)\rightarrow h(\pt,Zk^+)+q'(\kpt,(1-Z)k^+)$ is perfomed using the splitting function given in Eq. (\ref{eq:splitting explicit}).
In our simulation the splitting variable $Z$ is generated first. The differential probability is the integral of Eq. (\ref{eq:splitting explicit}) over $\kpt$ and writes
\begin{eqnarray}
\label{eq:Z distribution}
&&dZ\,\xi(Z)\,(1-Z)^a\,\exp\left(-\frac{\bl m_h^2}{Z}-\bt\,\xi(Z)\ktkt\right) \\
\nonumber &&\left[|\mu|^2+\frac{Z\xi(Z)}{\bl}+\xi(Z)^2\textbf{k}_{\rm T}^2-2\IM(\mu)\xi(Z)\,\textbf{S}_{int}\cdot\tilde{\textbf{k}}_{\rm T}\right].
\end{eqnarray}

The choice of the values of the parameters $a$, $\bl$, $\bt$ and $\mu$ entering Eq. (\ref{eq:Z distribution}) will be discussed in the next section. The terms which affect mostly the distribution of $Z$ are
\begin{itemize}
\item[-] the parameter $a$ which suppresses large values of $Z$ by a power law
\item[-] the exponential $\exp\left(-\bl\,m_h^2/Z-\bt\xi(Z)\ktkt\right)$ which depends on $m_h$ and on $\textbf{k}_{\rm T}^2$. This shifts $Z$ toward larger values when $\ktkt$ is large.
\end{itemize}
To be more precise, the first rank hadron $h_1$ is generated in the splitting $q_{\rm A}(\textbf{0}_{\rm T},k_{\rm A}^+)\rightarrow h_1(\textbf{p}_{1\rm T},Z_1k_{\rm A}^+)+q_2(\textbf{k}_{2\rm T},(1-Z_1)k_{\rm A}^+)$, hence $Z_1$ is drawn according to Eq. (\ref{eq:Z distribution}) with vanishing $\kt$ and only the mass $m_h$ enters the exponential. At the next step the hadron of rank two $h_2$ is generated in the splitting $q_2(\textbf{k}_{2\rm T},k_{2}^+)\rightarrow h_2(\textbf{p}_{2\rm T},Z_2k_2^+)+q_3(\textbf{k}_{3\rm T},(1-Z_2)k_2^+)$, therefore $Z_2$ is shifted towards larger values with respect to $Z_1$ because now in Eq. (\ref{eq:Z distribution}) enters a not vanishing $\textbf{k}^2_{2\rm T}$. At this point the splitting of $q_3$ is similar to that of $q_2$ and no differences are expected for hadrons of rank two or higher.

After the generation of $Z$ we draw $\pt$ according to the differential probability
\begin{eqnarray}
\label{eq:kt' distribution}
\nonumber && d^2\pt \exp{\left[-\frac{\bl}{Z\xi(Z)}\left(\kpt-\xi(Z)\kt\right)^2\right]}\\
&\times& \left[|\mu|^2+\kptkpt-2\IM(\mu)\textbf{S}_{int}\cdot\tilde{\textbf{k}'}_{\rm T}\right]
\end{eqnarray}
where $\kt$ is fixed from the previous hadron generation. For $\IM(\mu)>0$, the $\textbf{S}_{int}\cdot\tilde{\textbf{k}}'_{\rm T}$ term of Eq. (\ref{eq:kt' distribution}) pushes $\kpt$ in the direction of $\hat{\textbf{z}}\times\textbf{S}_{int}$ and contributes to the Collins effect through $\pt=\kt-\kpt$. This reproduces the classical $^3P_0$ mechanism. An other consequence is the spin-mediated correlation between $\kt$ and $\kpt$, which is negative for pseudoscalar meson emission. 
On the other hand, the exponential factor in Eq. (\ref{eq:kt' distribution}) naively forces the relation $\pt\sim (1-\xi(Z))\textbf{k}_{\rm T}$ and since $0<\xi(Z)<1$ for every value of $Z$, the effect is that the transverse momenta of two successive quarks tend to be aligned, as already mentioned. Hence in our model there are two effects at work, which are opposite for the case of pseudoscalar mesons: the ${}^{3}P_0$ mechanism and the dynamical correlations between the transverse momenta of quarks in the decay chain.
Since the $\pt$ distribution strongly depends on $\kt$ due to the exponential factor in Eq. (\ref{eq:kt' distribution}), we expect differences between the $\ptpt$ distributions of the first and second rank hadrons and no further change for higher rank hadrons.

\subsection{Values of the free parameters}
\label{sec:results}
The values of the four free parameters $a$, $\bl$, $\bt$ and $|\mu|^2$ have been tuned comparing the simulation results for unpolarized quark fragmentations with experimental data on multiplicities of charged hadrons in SIDIS off unpolarized deuteron as function of $p_{\rm T}^2$ \cite{compass-pt2} and with a set of unpolarized $\pt$-integrated fragmentation functions from global fits \cite{kniehl2001testing}, in order to find a satisfying qualitative agreement. The slope of the $p_{\rm T}^2$ spectrum is sensitive to $\bl$ and $\bt$ while its detailed shape for $p_{\rm T}^2\rightarrow 0$ is sensitive to $|\mu|^2$. The slopes are not affected by $a$, which changes the fragmentation functions at large hadron fractional energy $z_h$ and has been fixed comparing with the $\pt$-integrated FFs.
In this work we have used $a=0.9$, $\bl=0.5\,GeV^{-2}$, $\bt=5.17\,{(GeV/c)}^{-2}$ and $|\mu|^2=0.75\,(GeV^2/c^2)^2$.

The ratio $\IM(\mu)/\RE(\mu)$ has been fixed comparing the simulated and the measured Collins asymmetries extracted from $e^+e^-$ annihilation data, as explained in the next Section.
In all the simulations we use $\RE(\mu)=0.42\, GeV/c^2$, $\IM(\mu)=0.76\,GeV/c^2$.

Note that all the results but those in Sect. \ref{sec:kt} have been obtained with a vanishing primordial transverse momentum.


\subsection{Kinematical distributions}

\begin{figure*}[tb]
 \renewcommand{\thesubfigure}{a}
\begin{subfigure}[t]{.4\textwidth}
  \includegraphics[width=0.8\linewidth]{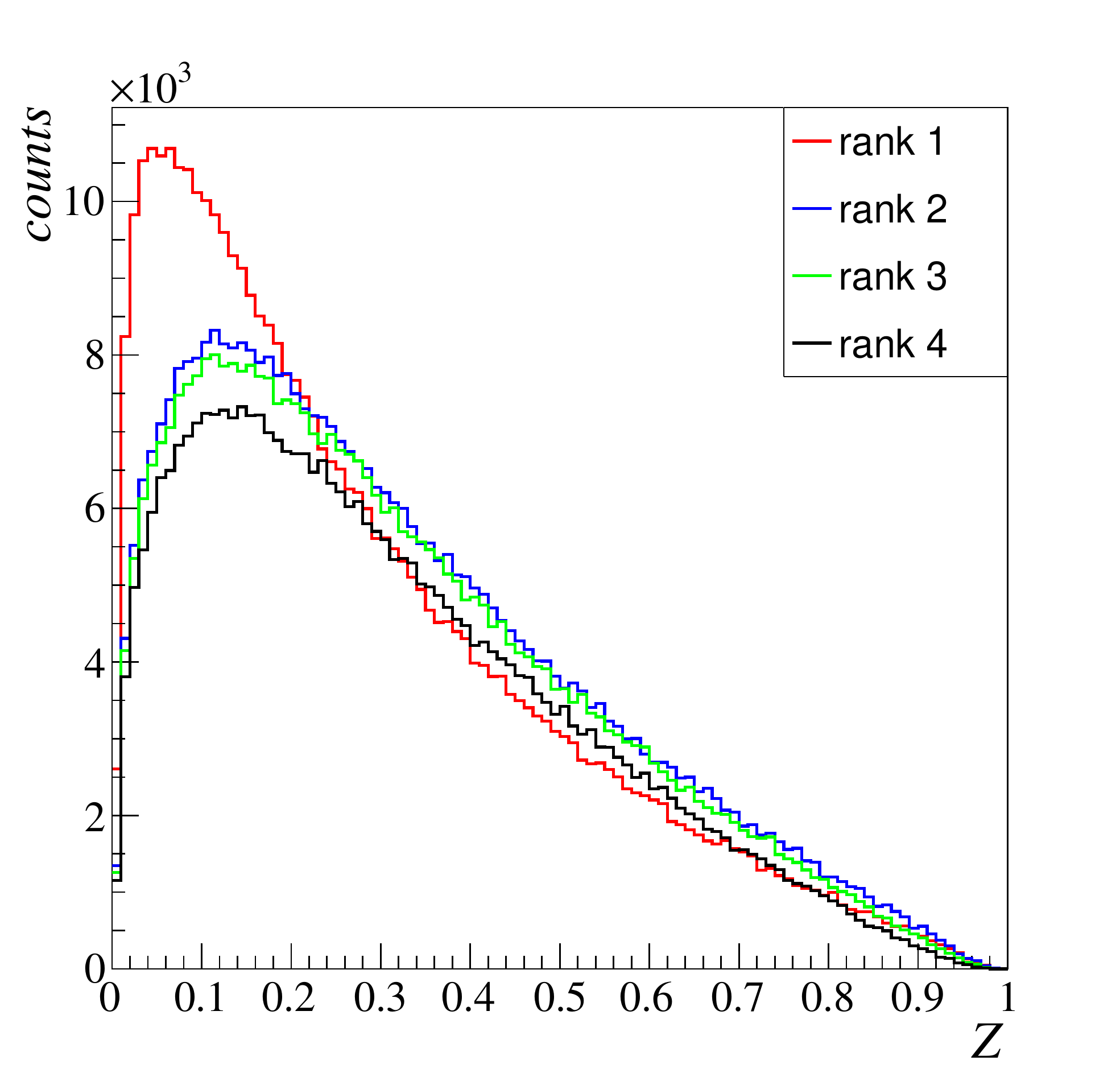}
 \caption{}
  \label{fig:Z distribution}
  \end{subfigure}
\renewcommand{\thesubfigure}{b}
\begin{subfigure}[t]{.4\textwidth}
	\includegraphics[width=0.8\linewidth]{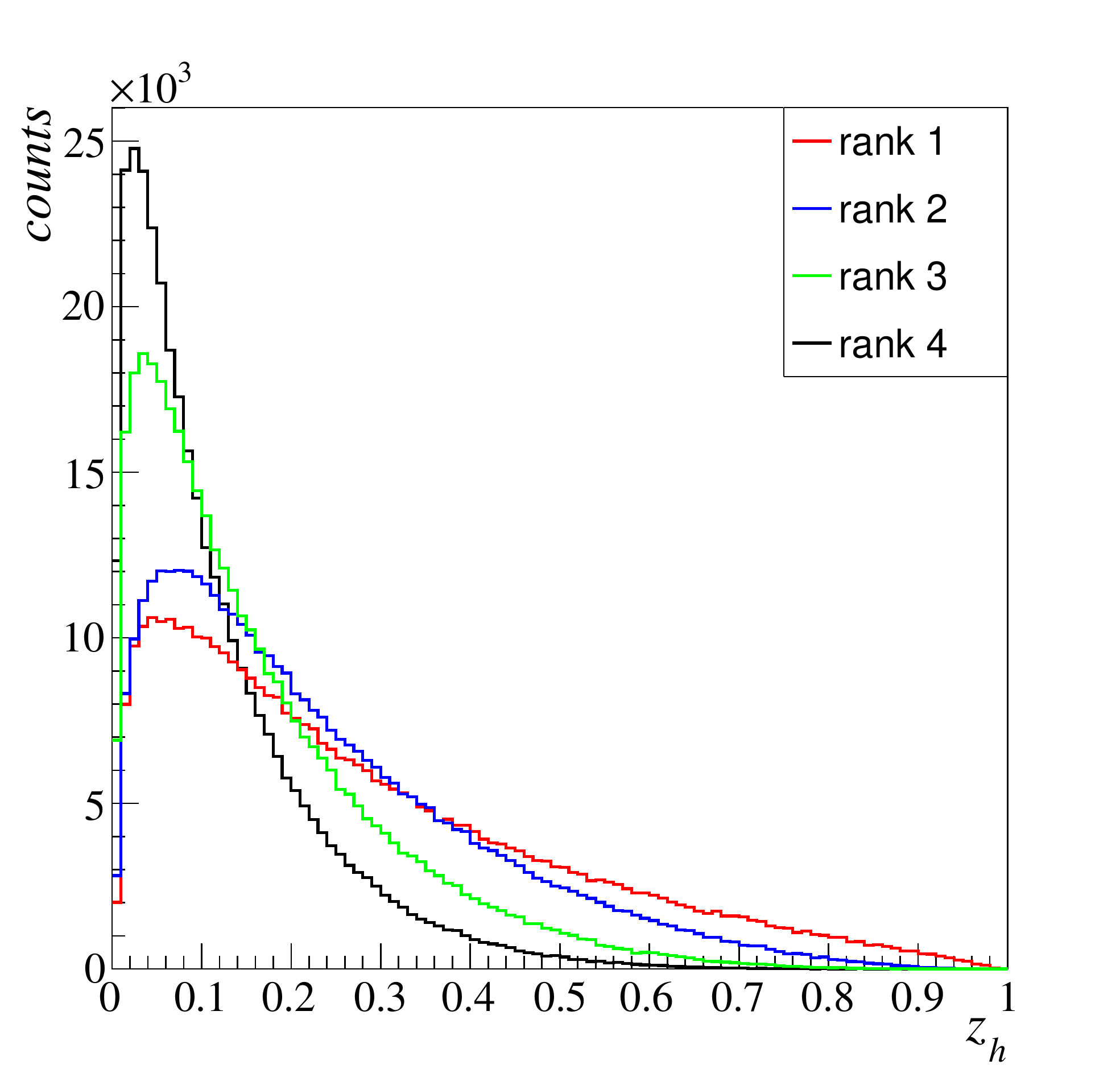}
	\caption{}
	\label{fig:zh distribution}
\end{subfigure}
\caption{Distributions of kinematical variables $Z$ (a) and $z_h$ (b) in the first four rank hadrons in $u$ quark jets.}
\end{figure*}

Let us first look at the kinematical (spin-independent) distributions with the chosen values of the parameters.
Figure \ref{fig:Z distribution} shows the distributions of the longitudinal splitting variable $Z$ for the first four rank hadrons generated in the fragmentation chain of a $u$ quark. As can be clearly seen, the distribution of the first rank hadron is shifted towards smaller values of $Z$ with respect to the distributions of higher rank hadrons. This is due to the $\kt$ dependence of the splitting function discussed above. Also, the distributions of higher rank hadrons are similar, as expected.

It is interesting to compare Fig. \ref{fig:Z distribution} with Fig. \ref{fig:zh distribution} showing the $z_h$ distributions for the first four rank hadrons. 
The shapes of the $z_h$ distributions are similar for rank $1$ and $2$ and change sensibly with the hadrons rank $r$ because of the relation $z_{h_r}\simeq Z_r(1-Z_{r-1})\cdot\cdot\cdot(1-Z_2)(1-Z_1)$. By definition the $Z$ and $z_h$ distributions for the rank one hadron coincide.

\begin{figure*}[tb]
 \renewcommand{\thesubfigure}{a}
\begin{subfigure}[t]{.4\textwidth}
  \includegraphics[width=0.8\linewidth]{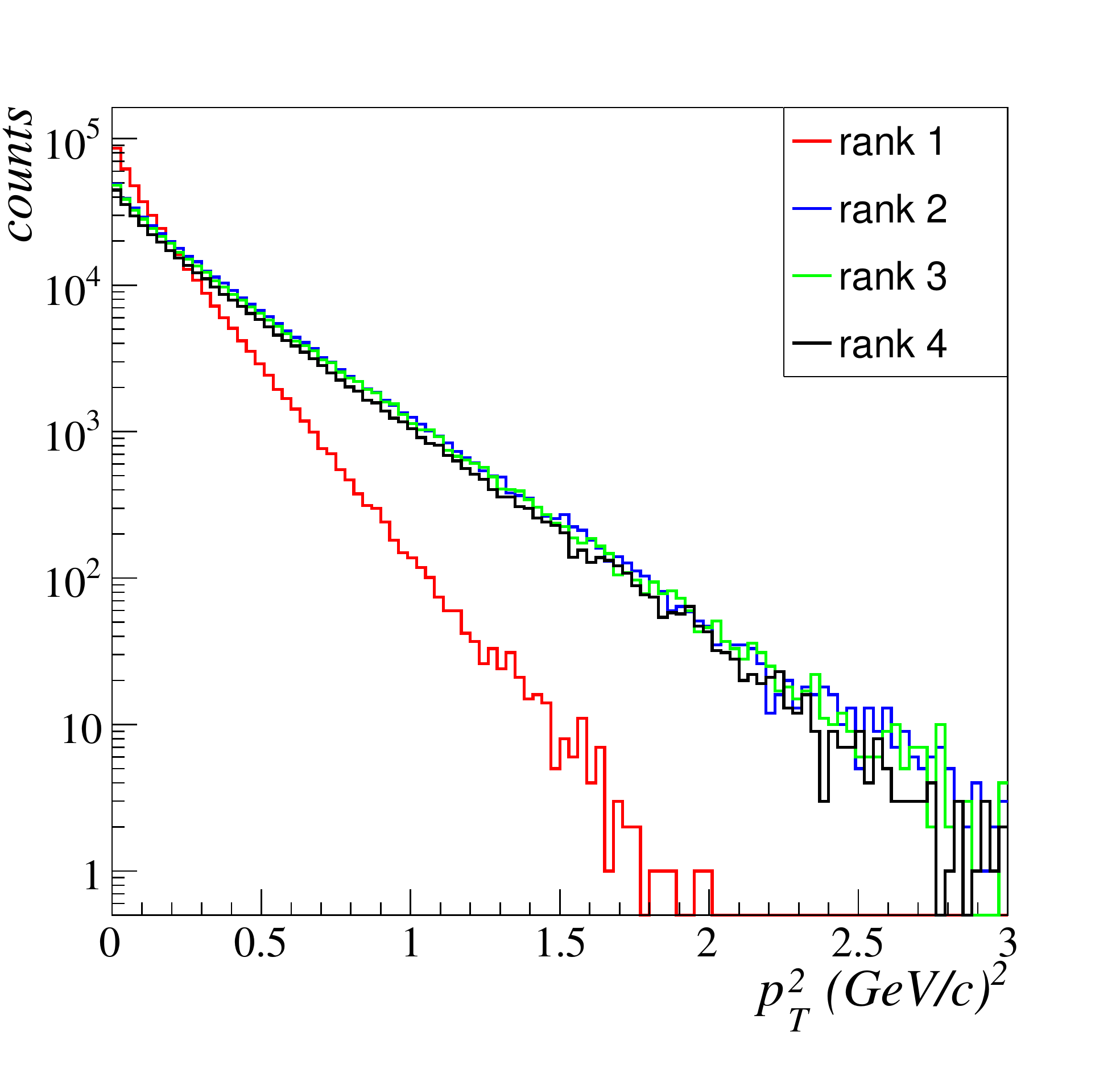}
 \caption{}
  \label{fig:pt distribution}
  \end{subfigure}
\renewcommand{\thesubfigure}{b}
\begin{subfigure}[t]{.4\textwidth}
  \includegraphics[width=0.8\linewidth]{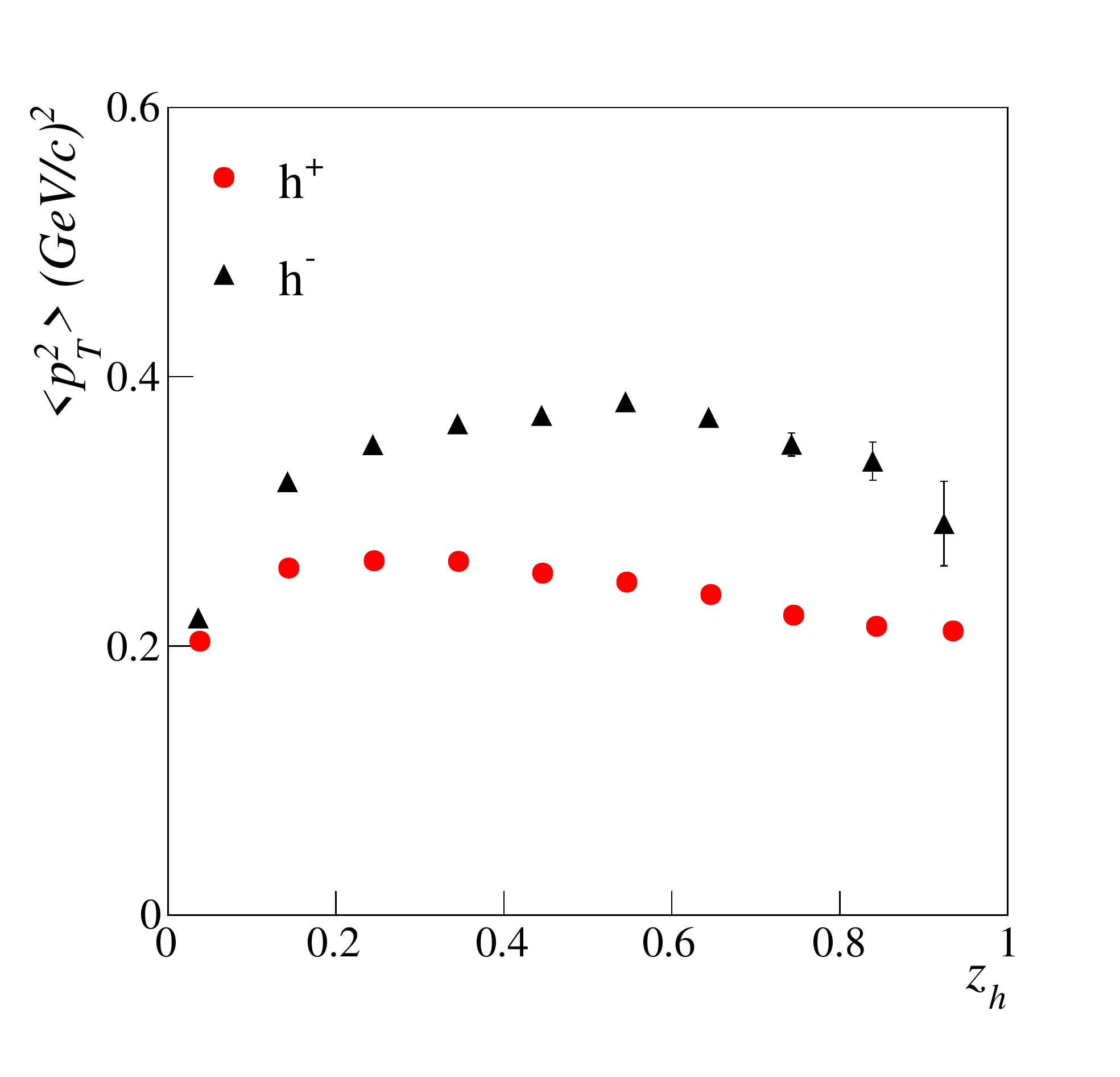}
 \caption{}
  \label{fig:pt2 zh}
  \end{subfigure}
  \caption{$\ptpt$ distribution of the first four rank hadrons (a) and $\langle \ptpt\rangle$ as function of $z_h$ (b) for positive and negative hadrons in $u$ quark jets.}
  \end{figure*}

The $\kptkpt$ distributions for the different splittings are very much the same and, since the initial quark has vanishing $\kt$, the $\kptkpt$ distribution of the left-over quark in the first splitting coincides with the $\ptpt$ distribution of the first rank hadron.
As a consequence the $\ptpt$ distribution of the first rank hadron is different from the distributions of higher rank hadrons, as shown in Fig. \ref{fig:pt distribution}. The slope of the $h_1$ distribution is almost twice the slope of $h_2$ distribution. This is due to the fact that $\langle \textbf{p}^2_{1 \rm T}\rangle=\langle \textbf{k}^2_{1 \rm T}\rangle$ whereas $\langle \textbf{p}^2_{2 \rm T}\rangle =\langle \textbf{k}^2_{2 \rm T}\rangle+\langle \textbf{k}^2_{3 \rm T}\rangle-2\langle \textbf{k}_{2 \rm T}\cdot\textbf{k}_{3 \rm T}\rangle\simeq 2\langle\textbf{k}^2_{1\rm T}\rangle$. This difference between the $\ptpt$ of rank $1$ and rank $2$ hadrons is a common feature of all recursive fragmentation models.
A trace of it is the inequality $\langle\ptpt(h^+)\rangle<\langle\ptpt(h^-)\rangle$ for medium values of $z_h$ in $u$ jets, shown in Fig. \ref{fig:pt2 zh}. However the data suggest the opposite. This discrepancy could be reduced with the introduction of resonances, like vector mesons, and their decay.
The decrease of $\langle\ptpt\rangle$ at small $z_h$ is due to the factor $\exp(-\bl\ptpt/Z)$ in the splitting function (Eq. \ref{eq: splitt spin}).

\section{Results on the transverse spin asymmetries}




In order to study the transverse spin effects measured in SIDIS off transversely polarized protons and in $e^+e^-$ annihilation, fragmentation events have been generated for initial quarks fully polarized along a fixed $\hat{\textbf{y}}$ axis ortogonal to the string axis. Only the results of the dominant $u$ quark fragmentation are shown in the following. We have checked that the results for pion production in $d$ quark fragmentation are related to those of the $u$ quark fragmentation by isospin symmetry.

For the SIDIS case we have used a sample of real COMPASS events. The $x_B$ and $Q^2$ of these events serve to fix the initial kinematics of our simulation event-by-event.


For the study of the asymmetries in the azimuthal distributions of the hadrons produced in $e^+e^-$ annihilation a second sample of events has been generated with a fixed initial quark energy corresponding to the BELLE energy.

In this section we present the results on the single hadron and the di-hadron transverse spin asymmetries and discuss the kinematical dependences of the corresponding analysing power. The effect of the primordial transverse momentum is also described. The Monte Carlo (MC) results are compared only with the COMPASS and BELLE data, which are in quite good agreement with the corresponding results from HERMES \cite{hermes} and Jefferson Lab experiments \cite{jlab} and from BaBar \cite{babar} and BESIII \cite{besIII} experiments respectively.

\subsection{Single hadron transverse spin asymmetries}


The well known Collins effect \cite{FFcollins} is the left-right asymmetry in the distribution of the hadrons produced in the fragmentation of a transversely polarized quark with respect to the plane defined by the spin and the momentum of the quark. The azimuthal distribution of the hadrons of the jet is given by
\begin{equation}
\frac{dN_h}{dz_h\,d^2\pt}\propto 1+a^{q\A\uparrow\rightarrow h+ X}(z_h,p_{\rm T})S_{\rm AT}\sin\phi_C
\end{equation}
where $S_{\rm AT}$ is the quark transverse polarization. The angle $\phi_C=\phi_h-\phi_{\textbf{S}_{\rm A}}$ is the Collins angle, where $\phi_h$ and $\phi_{S\A}$ are the azimuthal angles of the hadron momentum and of the quark spin. The analysing power $a^{q\A\uparrow \rightarrow h+ X}$ is the ratio between the spin dependent part of the FF (the Collins FF) and the unpolarized quark FF. Experimentally, the Collins effect has been observed in SIDIS, where the Collins FF couples with the transversity PDF, and in $e^+e^-$, where the measured azimuthal asymmetry can be written in terms of products of two Collins FFs.
In our model the same $\sin\phi_C$ is expected and no other azimuthal modulation is present.

Using simulated events, the analysing power $a^{q\A\uparrow\rightarrow h+ X}$ is calculated as $2\langle\sin\phi_C\rangle$ and in general it is function of both $z_h$ and $p_{\rm T}$. Since the model is formulated at the amplitude level it respects positivity. Indeed, from simulations we see that $|2\langle\sin\phi_C\rangle|<1$.

The Collins asymmetry measured from $e^+e^-$ annihilation data has been used to fix the value of the free parameter $\IM(\mu)=0.76\,GeV/c^2$. More specifically we have compared the mean value of the Collins analysing power for positive pions in transversely polarized $u$ jets from simulations with the mean value $0.258\pm 0.006$ obtained in Ref. \cite{M.B.B} from BELLE data.

\begin{figure*}[tb]\centering
\begin{minipage}{.7\textwidth}
  \includegraphics[width=0.9\textwidth]{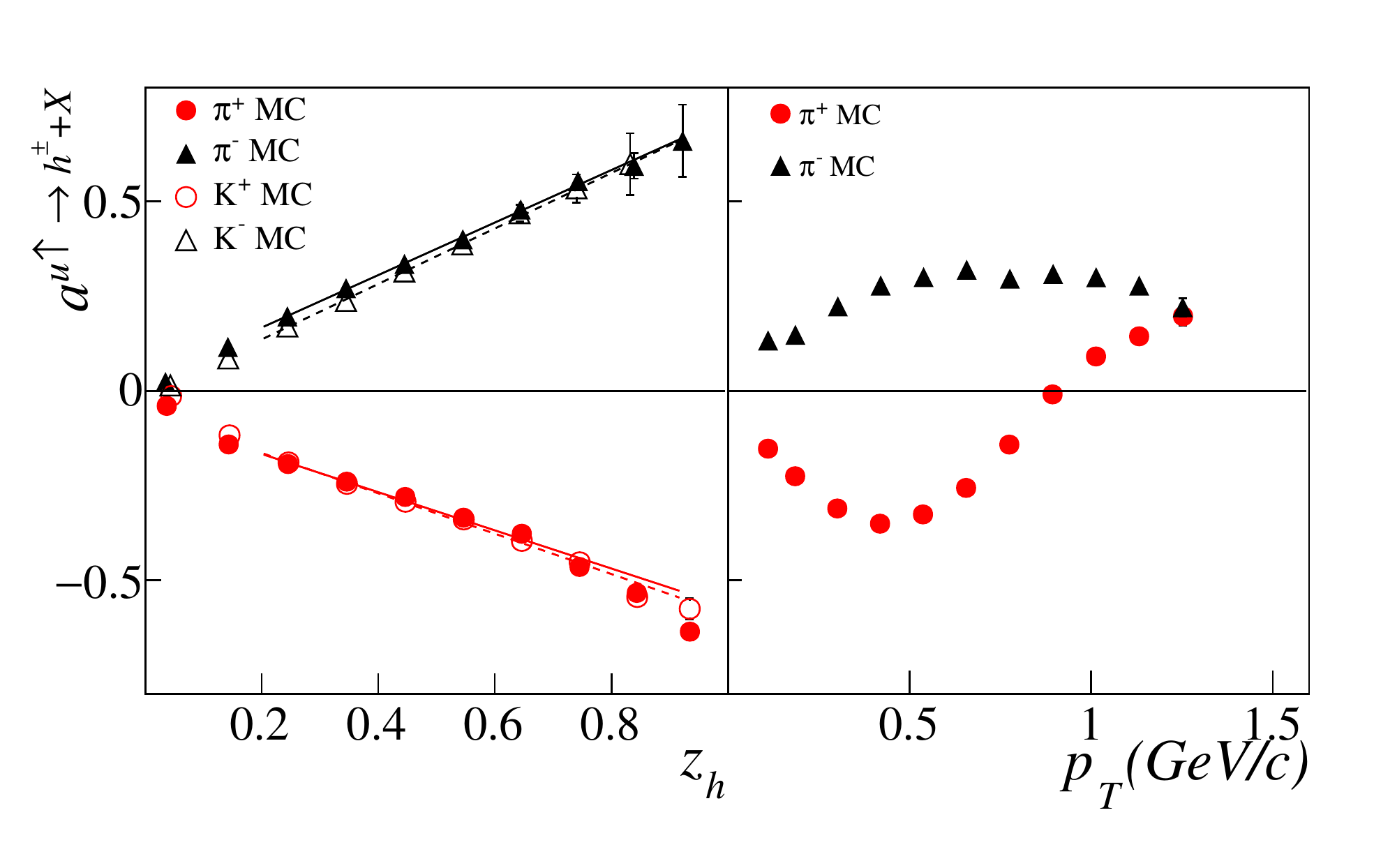}
\end{minipage}%
\caption{\small{Left panel, Collins analyzing power as function of $z_h$ for charged pions produced in simulations of transversely polarized $u$ quark jets. Right panel, simulated Collins asymmetry as function of $p_{\rm T}$. The cut $p_{\rm T}>0.1\,GeV/c$ is applied in both cases and $z_h>0.2$ only for the $p_{\rm T}$ analysing power.}}\label{fig:mc asymm}
\end{figure*}

Figure \ref{fig:mc asymm} shows the Collins analyzing power $a^{u\uparrow \rightarrow h + X}$ as function of $z_h$ for charged pions and kaons (left panel) and as function of $p_{\rm T}$ for charged pions (right panel). The main feature is that the analysing power has opposite sign and almost equal magnitude for oppositely charged mesons, as qualitatively expected from the ${}^3P_0$ model. The mean values for hadrons with $p_{\rm T}>0.1GeV/c$ and $z_h>0.2$ are given in Tab. \ref{tab:1h mean analyz power}.

\begin{table}[b]
\caption{\label{tab:1h mean analyz power}\small{Mean values of the analyzing power shown in Fig. \ref{fig:mc asymm} for positive and negative charges. The cuts $z_h>0.2$ and $p_{\rm T}>0.1\,GeV/c$ have been applied.}}
\begin{ruledtabular}
\begin{tabular}{l*{6}{c}r}
$\langle a^{u\uparrow\rightarrow h+X}\rangle$         & $h^+$ & $h^-$   \\
\hline
$\pi$& $-0.260\pm 0.002$ & $0.268\pm 0.002$   \\
$K$ & $-0.270\pm 0.003$ & $0.234\pm 0.004$\\
\end{tabular}
\end{ruledtabular}
\end{table}

 Also, the analyzing power vanishes for small $z_h$ and is almost linear in the range $0.2<z_h<0.8$.  A linear dependence on $z_h$ is also suggested by the BELLE data \cite{M.B.B} when the analysing power for the favoured fragmentation is assumed to be opposite to that for unfavoured fragmentation.

\begin{figure}[tb]\centering
\begin{minipage}{.4\textwidth}
\includegraphics[width=0.9\textwidth]{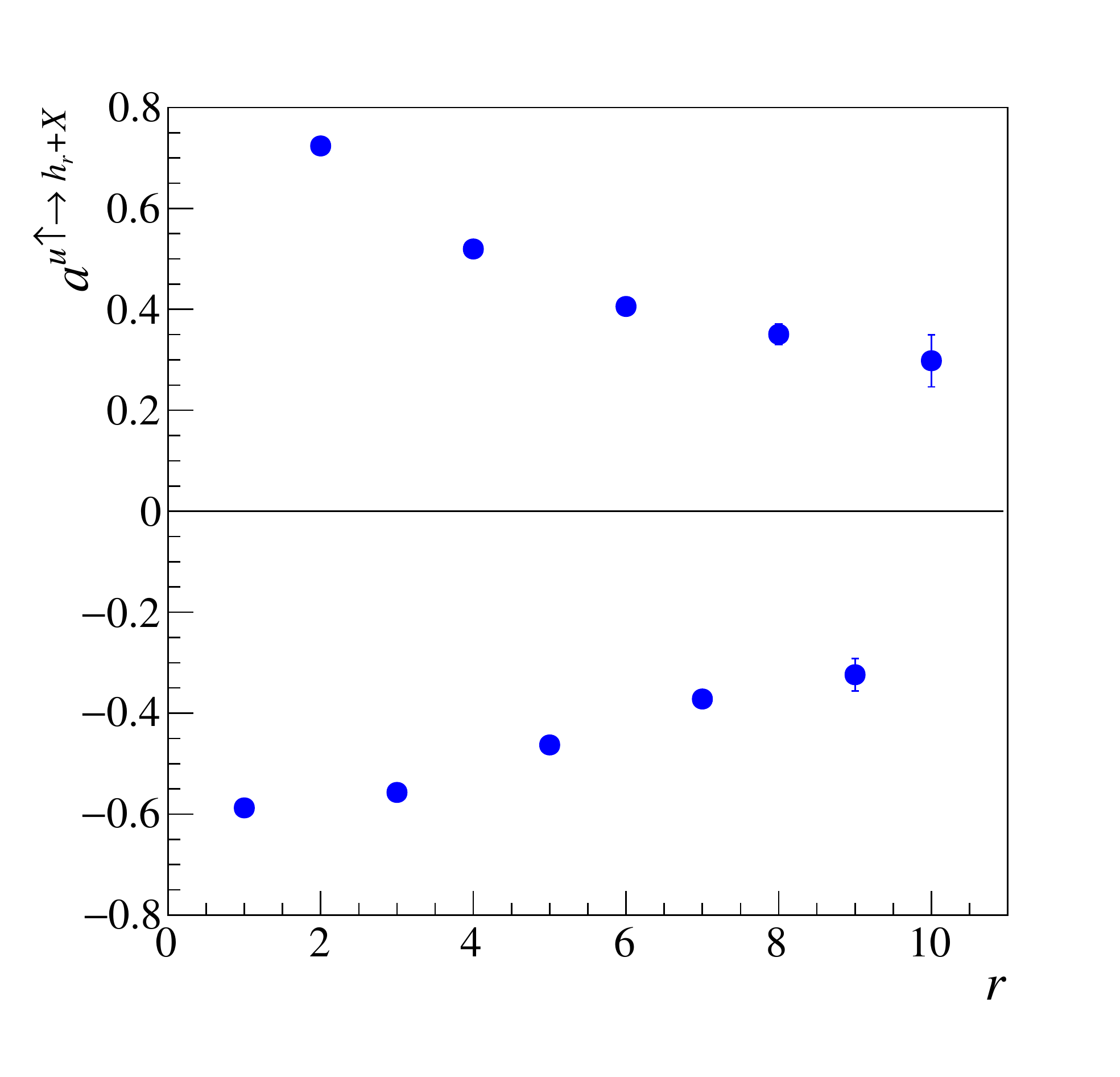}
\end{minipage}
\caption{Collins analysing power for positive pions as function of their rank. The cuts $z_h>0.1$ and $p_{\rm T}>0.1\,GeV/c$ have been applied.}\label{fig:ap rank}
\end{figure}

The sign and the simple dependence of the analysing power on $z_h$ can be understood performing a rank decomposition of the asymmetry by writing $a^{u\uparrow \rightarrow h + X}$ as the sum of different rank hadron contributions weighted by the number of hadrons of that rank. The analysing power can be written as
\begin{equation}
a^{u\uparrow \rightarrow h + X}(t)=\frac{\sum_r N_{h_r}(t)a^{u\uparrow\rightarrow h_r+X}(t)}{\sum_r N_{h_r}(t)}
\end{equation}
where the variable $"t"$ can be either $z_h$ or $p_{\rm T}$. $N_{h_r}$ is the number of hadrons of type $h$ and of rank $r$ and $a^{u\uparrow\rightarrow h_r + X}$ is the analysing power associated with rank $r$, both calculated at the same value $t$. The analysing power for the different rank hadrons is shown in Fig. \ref{fig:ap rank}. It has opposite sign for even and odd ranks, as a consequence of the local compensation of the quark transverse momenta and of the $^3P_0$ mechanism, and decreases with the rank. Such decrease is due to the depolarization of the recurrent quark which, with the current choice of parameters, turns out to be a weak effect. Indeed in each splitting roughly $10\%$ of the recurrent quark transverse polarization is lost.
The analysing power for a given hadron, being a mixture of different ranks, is then built as the combination of these effects.

Concerning the sign of the analysing power, for an initial $u$ quark, a fast positive pion can be produced at first rank or at rank $r>1$ following $r-1$ $\pi^0$'s or $\eta$'s. On the contrary a negative pion can never be produced at first rank because of its charge. Furthermore the contribution of larger ranks is smaller because $N_{h_r}(r)$ decreases with rank due to the finite $W$. Thus the sign of the $\pi^+$ and $\pi^-$ analysing powers is fixed by the contributions of the first and second ranks respectively. The same considerations can be made for charged kaons.

The almost linear dependence of the analysing power as function of $z_h$ is accidental: the relevant feature is the decay with $z_h$.

From the left panel of Fig. \ref{fig:mc asymm}, we notice also that the slope for negative mesons, which are unfavoured in $u$ chains, is slightly larger than the slope for positive ones. This effect is easily explained by the fact that the absolute value of the analysing power for a rank two hadron is somewhat larger than the analysing power for a rank one, as can be seen from Fig. \ref{fig:ap rank}.
Finally we can see that the slope for $\pi^-$ and $K^-$ are similar, as expected because both start to be produced from rank two.

Concerning the analysing power as function of $p_{\rm T}$, shown in the right panel of Fig. \ref{fig:mc asymm}, there are clearly different behaviours for positive and negative mesons.
An interesting feature is the change of sign of the analyzing power for positive pions at $p_{\rm T}\simeq 0.9\,GeV/c$.
The rank analysis at this value of $p_{\rm T}$ shows that the number of $\pi^+$ of rank $1$ and $3$ is roughly the same as the number of $\pi^+$ of rank $2$ and $4$. Moreover positive pions with large $p_{\rm T}$ are more likely produced as rank two, following a rank one $\pi^0$ or $\eta$, than as rank one. This effect combines with the alternate sign of the analysing power for even and odd rank hadrons to give $a^{u\uparrow\rightarrow \pi^+ + X}(p_{\rm T}=0.9\,GeV/c)\simeq 0$.
The number of higher rank pions decreases quickly and they give only a small contribution to the asymmetry.

\begin{figure*}[tb]\centering
\begin{minipage}{0.7\textwidth}
  \includegraphics[width=0.9\linewidth]{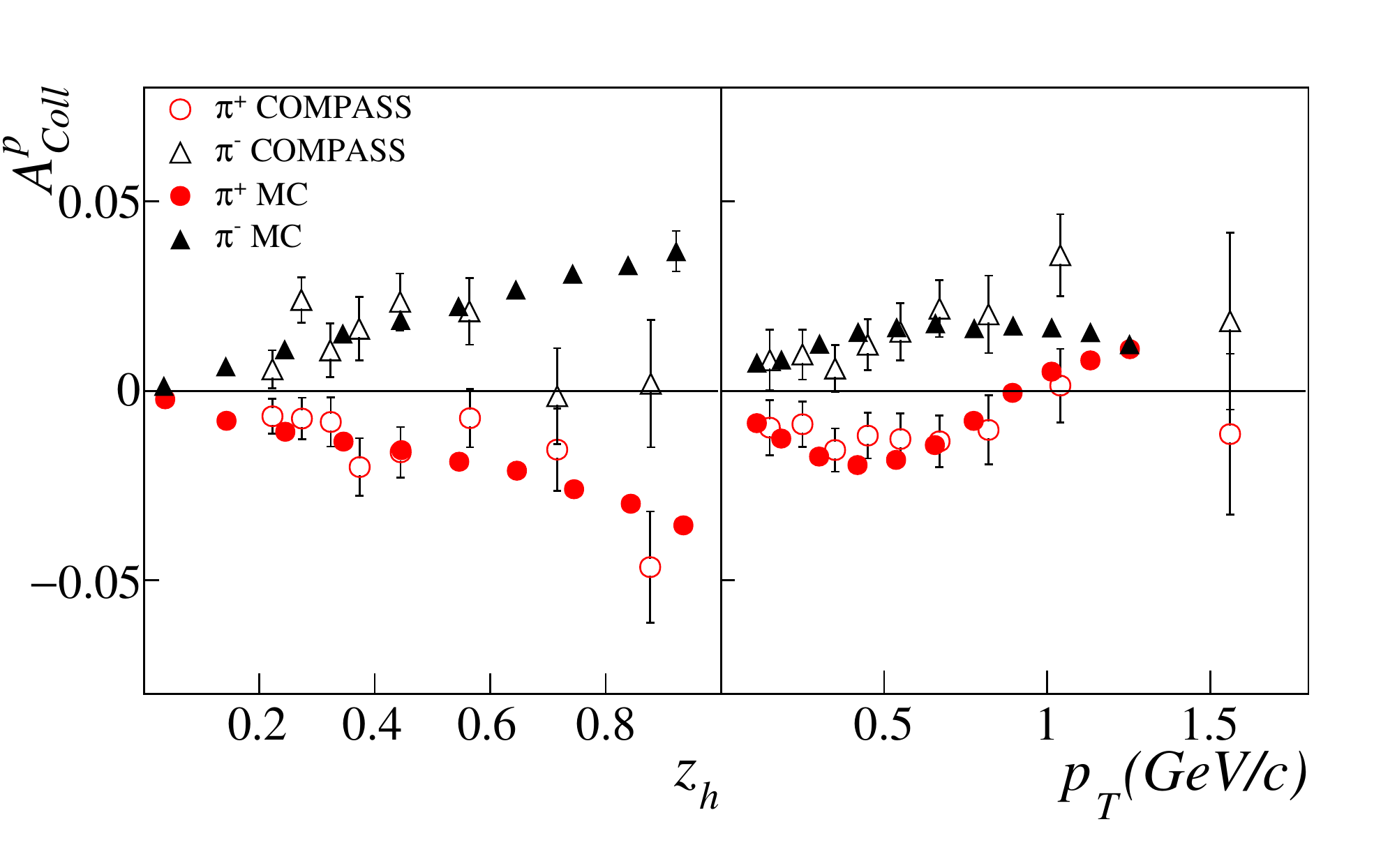}
\end{minipage}%
   \caption{\small{Comparison between the Collins asymmetry $A_{Coll}^p$ measured by COMPASS \cite{compassplb} (open points) and the Monte Carlo analysing power $a^{u\uparrow\rightarrow \pi^{\pm}+X}$ scaled by $\lambda$ (full points): as function of $z_h$ (left panel) and of $p_{\rm T}$ (right panel). The cuts $p_{\rm T}>0.1\,GeV/c$ and $z_h>0.2$ have been applied in both cases.}}\label{fig:comp asymm}
\end{figure*}

Similar trends are observed in the Collins asymmetry for charged pions produced in SIDIS off transversely polarized protons as measured by COMPASS \cite{compassplb}. The comparison with Monte Carlo results is shown in Fig. \ref{fig:comp asymm} as function of $z_h$ (left plot) and as function of $p_{\rm T}$ (right plot).
The Monte Carlo values in both panels are those of Fig. \ref{fig:mc asymm} multiplied by an overall scale factor $\lambda_{1}$ obtained from $\chi^2$ minimization. In the $u$-dominance hypothesis for a proton target and neglecting the primordial transverse momentum, $\lambda_1$ is the ratio of the $x_B$-integrated $u$-quark transversity and the $x_B$-integrated unpolarized $u$ quark density, multiplied by the depolarization factor of lepton-quark scattering.

As apparent from the right panel of Fig. \ref{fig:comp asymm}, the Monte Carlo describes qualitatively the $p_{\rm T}$ dependence of the experimental points, which do not exclude a change of the $\pi^+$ asymmetry sign for $p_{\rm T}>0.9\,GeV/c$.


The agreement between Monte Carlo and COMPASS asymmetries as function of $z_h$ is satisfactory for positive pions, whereas for negative pions it is poor for $z_h>0.6$. In this region, however, the contributions of $d$ quark fragmentation or of $\rho^0$ decay could be not negligible. 

\subsection{Dihadron transverse spin asymmetries}
The properties of the analyzing power $a^{u\uparrow \rightarrow h_1h_2+X}$ due to the Collins effect in the $h_1h_2$ pair production in a $u$ jet have also been studied. Such analyzing power has been found to be related to $a^{u\uparrow \rightarrow h^{\pm}+X}$ in a recent experimental work in SIDIS \cite{interplay} and its magnitude can be obtained by $e^+e^-$ data \cite{belle}.


In general the distribution of oppositely charged hadron pairs in the same jet, as function of the relevant variables used here, is given by
\begin{equation}
\frac{dN_{h_1h_2}}{dz\,dM_{inv}\,d\Phi}\propto 1+a^{q\A\uparrow\rightarrow h_1h_2+X}(z,M_{inv})S_{\rm AT}\sin(\Phi-\phi_{\textbf{S}_{\rm A}})
\end{equation}
where $z=z_{h_1}+z_{h_2}$, $M_{inv}$ is the invariant mass of the $h_1h_2$ pair and $\Phi$ is the azimuthal angle of a vector characterizing the pair (different choices have been made in different analysis). The subscript $1(2)$ indicates the positive (negative) hadron with transverse momentum $\textbf{p}_{1 \rm T}(\textbf{p}_{2 \rm T})$.

The analysing power is extracted from the simulated events as $2\langle \sin(\Phi-\phi_{S\A})\rangle$ taking into account all possible pairs of the jets.

\smallskip
\ni{\it Comparison with BELLE data.}
In order to compare with the $e^+e^-$ data we have evaluated the quantity $\epsilon(M_{inv})\equiv\langle a^{u\uparrow\rightarrow \pi^+\pi^-+X}\rangle a^{u\uparrow\rightarrow \pi^+\pi^-+X}(M_{inv})$, where $\langle a^{u\uparrow\rightarrow \pi^+\pi^-+X}\rangle$ is the analyzing power averaged over all the kinematical variables, including $M_{inv}$. For this comparison, in the simulation the analyzing power $a^{u\uparrow\rightarrow \pi^+\pi^-+X}$ has been estimated using $\Phi=\phi_{\rm BELLE}$ where $\phi_{\rm BELLE}$ is the azimuthal angle of the vector $\textbf{p}_{1\rm T}-\textbf{p}_{2\rm T}$.
 
\begin{figure}[tb]\centering
\begin{minipage}{.4\textwidth}
  \includegraphics[width=0.9\textwidth]{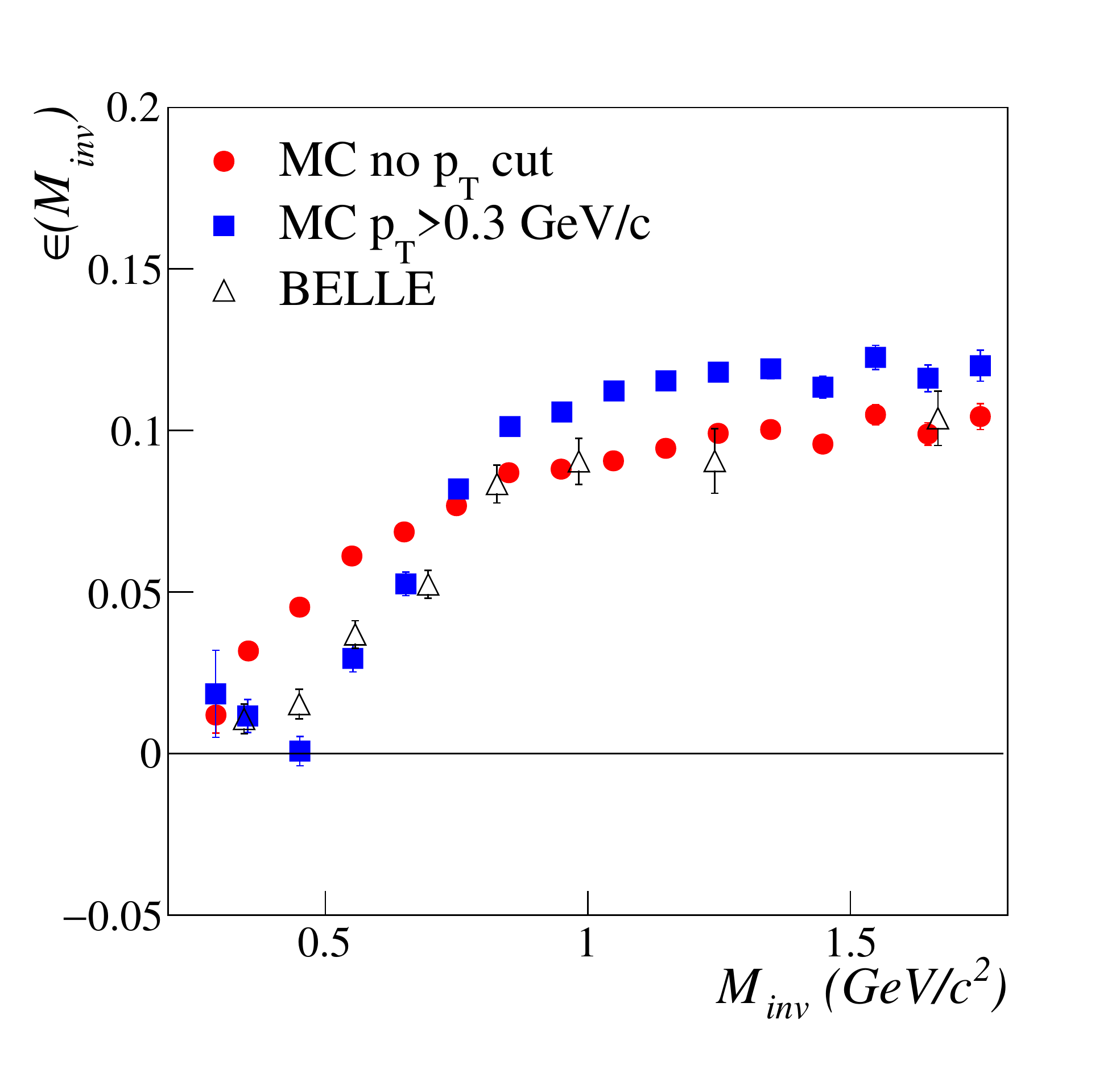}
\end{minipage} 
\caption{\small{Monte Carlo calculation of $\epsilon(M_{inv})$ for pions pairs produced in transversely polarized $u$ jets asking for each pion of the pair $z_h>0.1$ (circles) and also $p_{\rm T}>0.3\, GeV/c$ (squares). The black open triangles are the values of $\epsilon(M_{inv})$ obtained from BELLE data \cite{belle}.}}\label{fig:minv pt2>0.1}
\end{figure}

Figure \ref{fig:minv pt2>0.1} shows the results for $\epsilon(M_{inv})$ from the simulation when $z_{h1,2}>0.1$ with no cut in $p_{\rm T}$ as circles whereas for those represented by squares we have required $p_{\rm T}>0.3\,GeV/c$. The open triangles show the values of $\epsilon$ as measured by BELLE \cite{belle}.
Both in the simulation and in the data the analyzing power shows a saturation for large values of the invariant mass while for small values it tends to zero.
It has to be noted that the Monte Carlo data sample has a different invariant mass spectrum with respect to BELLE data. In particular in the BELLE data sample the statistics is larger in the region of the $\rho$ meson while in the program there are no resonances and most of the statistics is at higher values of $M_{inv}$. Still, both in BELLE and in simulation results, no structure can be seen.

We recall that, in order to cancel, or minimize, the effects due to the primordial transverse momenta, the dihadron asymmetry is normally written in terms of the azimuth of the relative transverse momentum
\begin{equation}\label{eq:RT}
 \textbf{R}_{\rm T}=(z_{h_2}\textbf{p}_{1\rm T}-z_{h_1}\textbf{p}_{2\rm T})/z.
 \end{equation}
 For the BELLE results we are considering here, the vector characterizing the pair is
\begin{equation}\label{eq:PT}
\textbf{p}_{1\rm T}-\textbf{p}_{2\rm T}=2\textbf{R}_{\rm T}+(z_{h_1}-z_{h_2})\textbf{P}_{\rm T}/z
\end{equation}
 where $\textbf{P}_{\rm T}=\textbf{p}_{1\rm T}+\textbf{p}_{2\rm T}$ is the global transverse momentum of the pair. Defining as "pure" di-hadron asymmetry the one defined with respect to the vector $\textbf{R}_{\rm T}$, the asymmetry extracted from the BELLE data is a combination of the "pure" di-hadron asymmetry and of the global Collins effect of the pair.

\begin{figure*}[tb]
\begin{minipage}{.7\textwidth}
  \includegraphics[width=0.9\linewidth]{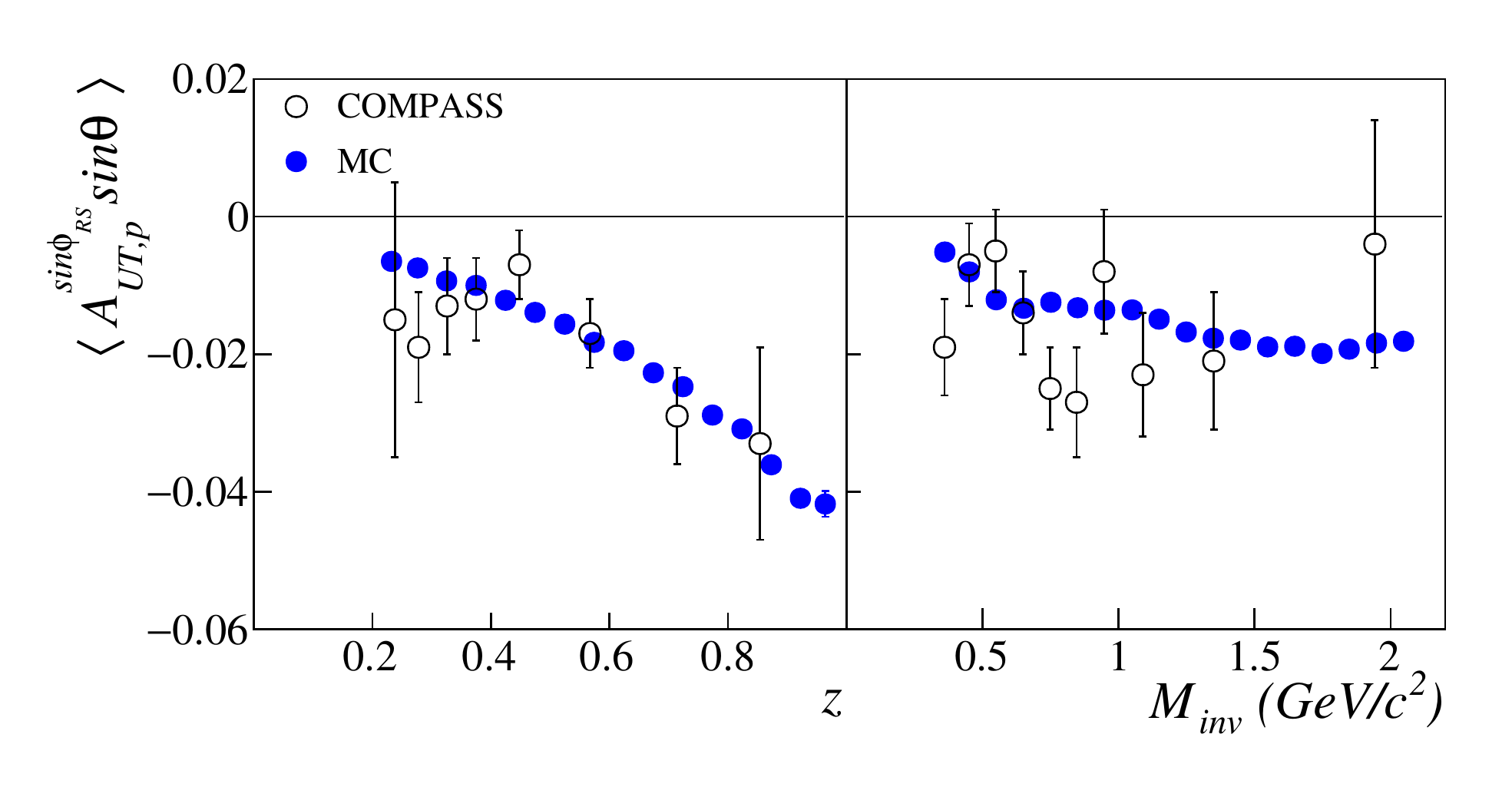}
\end{minipage}
 \caption{\small{Comparison between the di-hadron asymmetry $\langle A_{\rm UT,p}^{\rm \sin\phi_{\rm RS}}\rm \sin\theta\rangle$ (open points) as measured in COMPASS \cite{compass-dihadron} and the di-hadron analysing power calculated from the Monte Carlo (full points) as function of $z=z_{h_1}+z_{h_2}$ (left plot) and as function of $M_{inv}$ (right plot)}.}
  \label{fig:zh compass}
\end{figure*}

\smallskip
\ni{\it Comparison with COMPASS data.}
In Fig. \ref{fig:zh compass} we show the comparison between the Monte Carlo and the COMPASS dihadron asymmetry for $h^+h^-$ pairs measured in SIDIS off transversely polarized protons as function of $z$ (left) and $M_{inv}$ (right). The dihadron asymmetry is extracted using $\Phi=\phi_R$ where $\phi_R$ is the azimuthal angle of the vector $\textbf{R}_{\rm T}$, thus it can be regarded as a pure di-hadron asymmetry. Both in COMPASS data and in simulations the cuts $z_{h}>0.1$, $x_F>0.1$, $R_{\rm T}>0.07\,GeV/c$ and $|\textbf{p}_i|>3\,GeV$ ($i=1,2$) have been applied. 

The left plot of Fig. \ref{fig:zh compass} concerns the dependence on $z$. The Monte Carlo points are scaled by a factor $\lambda_2$ estimated by comparing with the COMPASS asymmetry as function of $z$. From a $\chi^2$ minimization we obtain $\lambda_2=0.055\pm 0.008$ in perfect agreement with the value of $\lambda_1$ obtained in the single hadron asymmetry case, as expected. The results from the Monte Carlo are in good agreement with the experimental data.

The right plot of Fig. \ref{fig:zh compass} shows the dependence of the analysing power on $M_{inv}$. The same cuts as those for the dihadron asymmetry as function of $z$ have been applied. After scaling by the same parameter $\lambda_2$, the Monte Carlo points describe quite well the trend of the data.

\subsection{Comparison between single hadron and dihadron transverse spin asymmetries}
Following the work done in Ref. \cite{interplay} we have studied the relationship between the Collins and the di-hadron analyzing powers for hadron pairs in the same $u$ quark jet, as function of the relative azimuthal angle $\Delta\phi=\phi_1-\phi_2$. In that analysis using only the events with at least one $h^+$ and one $h^-$ two kinds of asymmetries had been extracted: the "Collins Like" (CL) asymmetries $A_{CL1(2)}^{\sin\phi_C}$ for positive and negative hadrons and the dihadron asymmetry for oppositely charged hadron pairs $A_{CL,2h}^{\sin\phi_{2h,S}}$. In each bin of $\Delta\phi$, the CL asymmetry is the Collins asymmetry of $h^+$ ($h^-$) of the pair.

As in Ref. \cite{interplay} we calculate $a^{u\uparrow\rightarrow h^+h^-+X}$ using $\Phi=\phi_{2h}$, where $\phi_{2h}$ is the azimuthal angle of the vector $\hat{\textbf{p}}_{1\rm T}-\hat{\textbf{p}}_{2\rm T}$ and $\hat{\textbf{p}}_{\rm T}\equiv \textbf{p}_{\rm T}/|\textbf{p}_{\rm T}|$. 
Due to the relation
\begin{eqnarray}\label{eq:pT compass}
\nonumber \hat{\textbf{p}}_{1\rm T}-\hat{\textbf{p}}_{2\rm T}&=&\textbf{R}_{\rm T}(1/|\textbf{p}_{1\rm T}|+1/|\textbf{p}_{2\rm T}|)\\
&+&\textbf{P}_{\rm T}\frac{z_{h_1}/|\textbf{p}_{1\rm T}|-z_{h_2}/|\textbf{p}_{2\rm T}|}{z},
\end{eqnarray}
the considered asymmetry is a combination of the "pure" dihadron asymmetry and of the global Collins asymmetry of the hadron pair. However, as discussed in Ref. \cite{interplay}, the azimuthal angle $\phi_R$ is strongly correlated with $\phi_{2h}$, and the dihadron asymmetry measured from $2\langle \sin\phi_{2h,S}\rangle$ with $\phi_{2h,S}=\phi_{2h}-\phi_{S\A}$, is essentially the same as the "pure" dihadron asymmetry, which could be verified with the code as well.

\begin{figure*}[tb]\centering
 \renewcommand{\thesubfigure}{a}
\begin{subfigure}[b]{.4\textwidth}
  \includegraphics[width=0.9\textwidth]{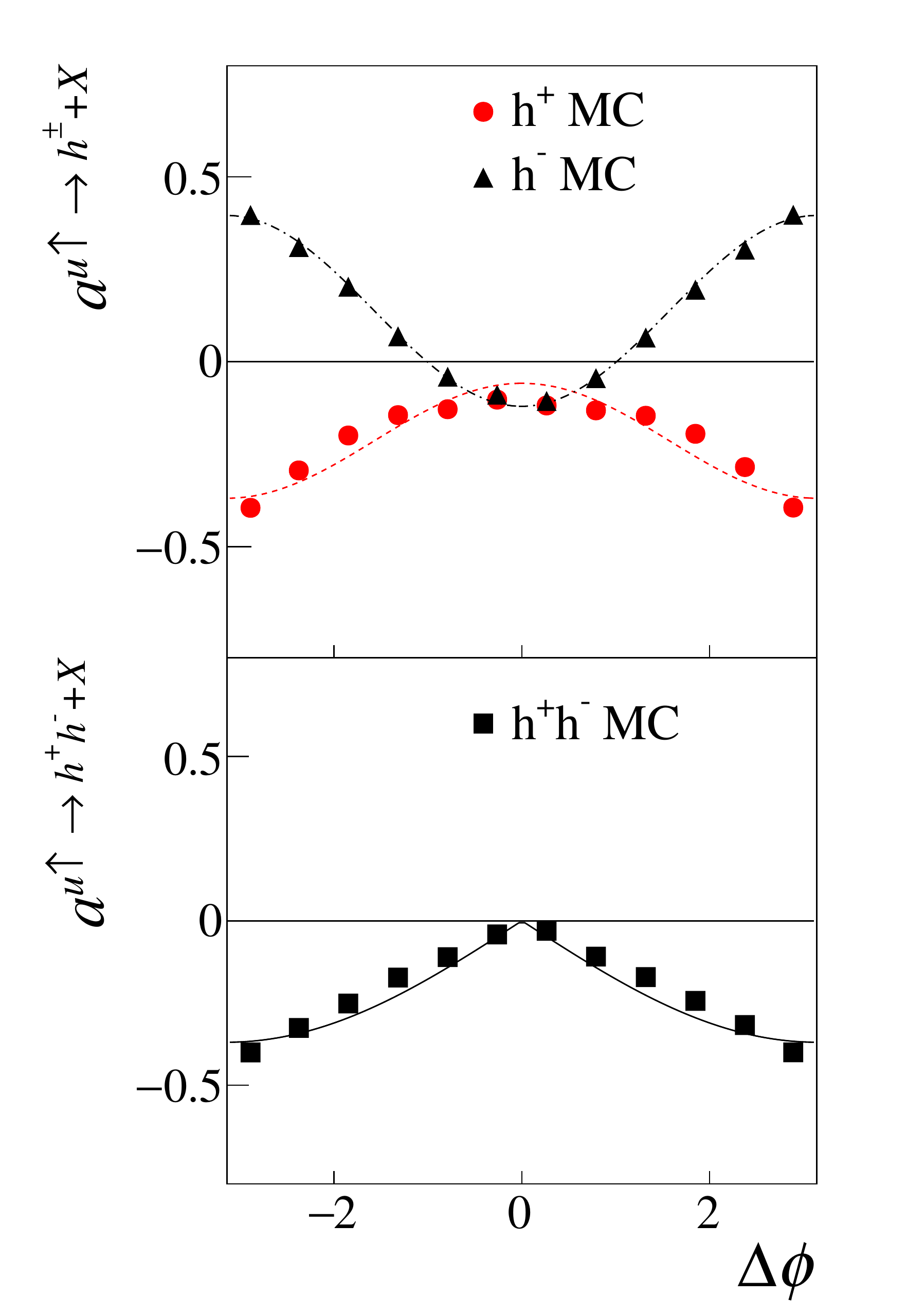}
  \caption{}
\end{subfigure}%
 \renewcommand{\thesubfigure}{b}
\begin{subfigure}[b]{.4\textwidth}
  \includegraphics[width=0.9\textwidth]{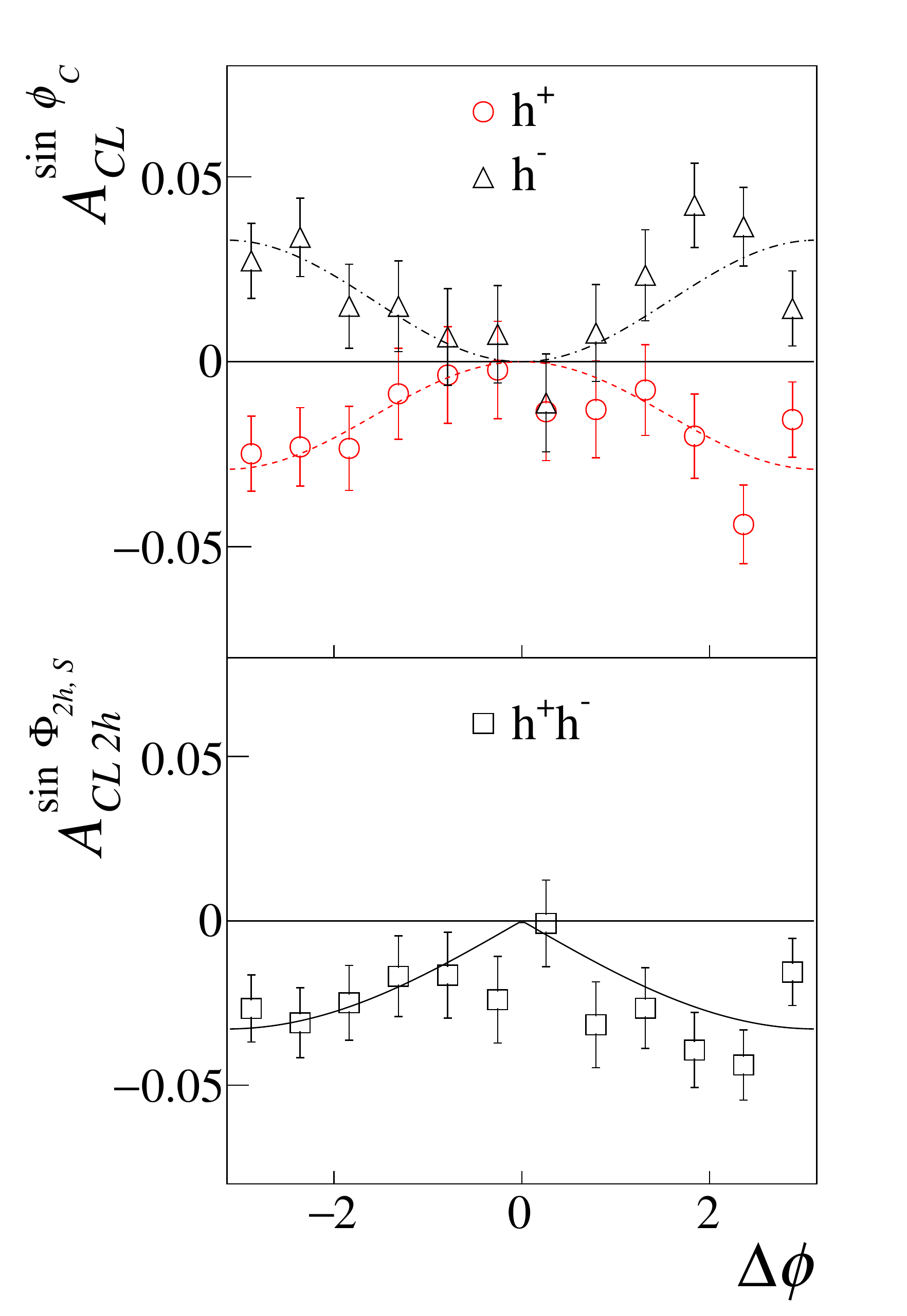}
  \caption{}
\end{subfigure}%

\caption{\small{(a): Monte Carlo results for the analyzing power in the case of "Collins Like" asymmetries (top) and di-hadron asymmetries (bottom) as function of $\Delta\phi$. (b): the corresponding asymmetries measured by COMPASS \cite{interplay}.}}\label{fig:delta phi}
\end{figure*}

The blue squares in Fig.\ref{fig:delta phi} (a) represent the di-hadron analyzing power $a^{u\uparrow\rightarrow \pi^+\pi^-+X}$ calculated in the Monte Carlo as function of $\Delta\phi$. The blue curve is the result of the fit with the function $c\sqrt{2(1-\cos\Delta\phi)}$ as suggested in Ref. \cite{interplay}.
The plot in Fig. \ref{fig:delta phi} (b) shows the asymmetry $A_{CL,2h}^{\sin\phi_{2h,S}}$ as measured in COMPASS. As can be seen, the agreement is good and in particular the mirror symmetry between $h^+$ and $h^-$ is clear in both cases. Note that the $A_{CL,2h}^{\sin\phi_{2h,S}}$ asymmetry is smaller than $a^{u\uparrow\rightarrow \pi^+\pi^-+X}$ by a factor of $0.1$ analogous to $\lambda_2$ but for the higher cut $x_B>0.032$ adopted in this experimental analysis.

The same considerations hold also for CL analysing power $A_{CL1(2)}^{\sin\phi_C}$ of $h^+$ and $h^-$ shown in the top plot of Fig.\ref{fig:delta phi} (a) with red circles and black triangles respectively. The corresponding COMPASS data are shown in top plot of Fig. \ref{fig:delta phi} (b): again, the trend is very similar. The MC points are fitted with functions of the type $\delta_{1(2)}+c_{1(2)}\sin\Delta\phi$, as suggested from Ref. \cite{interplay}, and the results are represented by the red and the black dashed lines.
The slight up-down disymmetry for $h^+$ and $h^-$ in the simulated results is due to the different values of the analyzing power for $h^+$ and $h^-$. The red and the black dashed lines in Fig.\ref{fig:delta phi} (b) represent the fits to the experimental CL asymmetries as shown in Ref. \cite{interplay}, which are consistent with vanishing $\delta_{1(2)}$ parameters.

As a conclusion, the Collins asymmetry and the dihadron asymmetry are generated by the same physical mechanism, which in our case is the string $+^3P_0$ hypothesis.

\subsection{Introducing the primordial transverse momentum}\label{sec:kt}

\begin{figure}[tb]
\begin{minipage}{.4\textwidth}
  \includegraphics[width=0.9\linewidth]{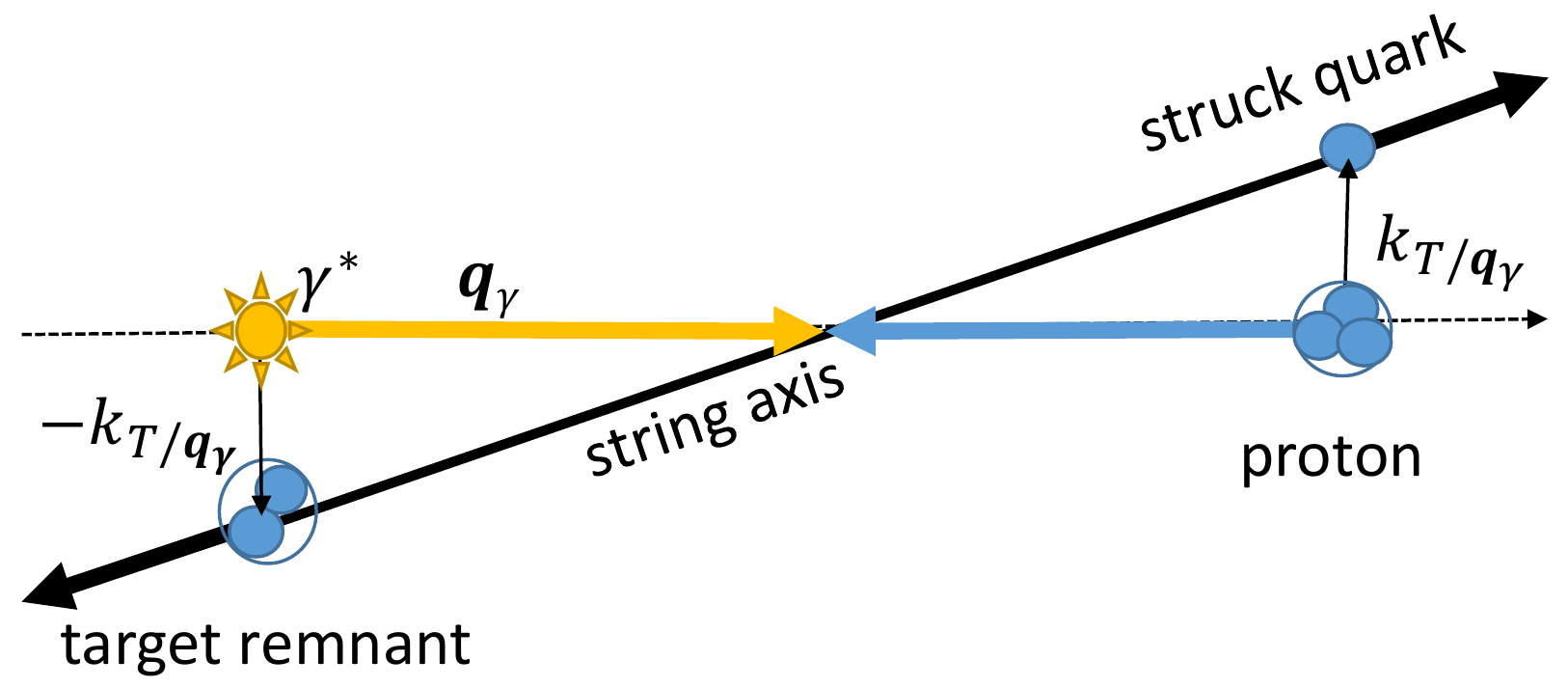}
  \end{minipage}
 \caption{\small{Illustration of the rotation of the string axis in the string center of mass frame.}}
  \label{fig:rotation}
\end{figure}

In the previous sections we did not consider the primordial transverse momentum of the initial quark. In this section we show the results when the initial quark $q_A$ does have a primordial transverse momentum. Figure \ref{fig:rotation} depicts the string direction in the DIS $\gamma^*$-nucleon center of mass frame when the struck quark has primordial transverse momentum $\textbf{k}_{\rm T prim}$, inherited from the quark motion in the nucleon.  $\textbf{k}_{\rm T prim}$, also written $\textbf{k}_{\rm T /\textbf{q}_{\gamma}}$, is defined with respect to the  $\gamma^*$ momentum $\textbf{q}_{\gamma}$. The target remnant has the opposite $-\textbf{k}_{\rm T prim}$. The string is stretched between $q_A$ and the target remnant. Its axis is therefore rotated from the $\gamma^*$-nucleon axis.
The effect of a random $\textbf{k}_{\rm T prim}$ is the broadening of the spectra of hadrons transverse momenta with respect to $\textbf{q}_{\gamma}$. This should partly smear the single hadron asymmetry.

The primordial momentum $\bf{k}_{\rm T prim}$ is generated according to the gaussian distribution
\begin{equation}
d^2\textbf{k}_{\rm T prim}\, \pi \langle k_{\rm T prim}^2\rangle^{-1}\,\exp(-\bf{k}_{\rm T prim}^2/\langle k_{\rm T prim}^2\rangle)
\end{equation}
where $\langle k_{\rm T prim}^2\rangle$ is a free parameter.
The fragmentation of the initial transversely polarized quark $q_A$ is performed using the rotated string axis as $\hat{\textbf{z}}$ axis. Then we go in the laboratory frame with a boost along the $\gamma^*$-nucleon axis. 

In the small angle approximation, the rotation in the string center of mass frame is practically equivalent to make the following shift in $\pt$ (which is relative to the string axis)
\begin{equation}\label{eq:shift}
\textbf{p}_{\rm T/\textbf{q}_{\gamma}}=x_F\,\textbf{k}_{\rm T prim}+\pt
\end{equation}
where $\textbf{p}_{\rm T/\textbf{q}_{\gamma}}$ is the hadron transverse momentum with respect to the $\gamma^*$ axis and $x_F=(2p_z/W)_{c.m.}$ is the Feynman scaling variable. Since $x_F=z_h-\epsilon_h^2/(z_hW^2)$, Eq. (\ref{eq:shift}) almost coincides for large $x_F$ with the often used equation $\textbf{p}_{\rm T/\textbf{q}_{\gamma}}=z_h\,\textbf{k}_{\rm T prim}+\pt$. The shift is zero at $x_F=0$ and opposite to $\textbf{k}_{\rm T  prim}$ in the backward hemisphere as can be guessed from Fig. \ref{fig:rotation}.

\begin{figure}[tb]\centering
\begin{minipage}{.4\textwidth}
 \includegraphics[width=0.9\linewidth]{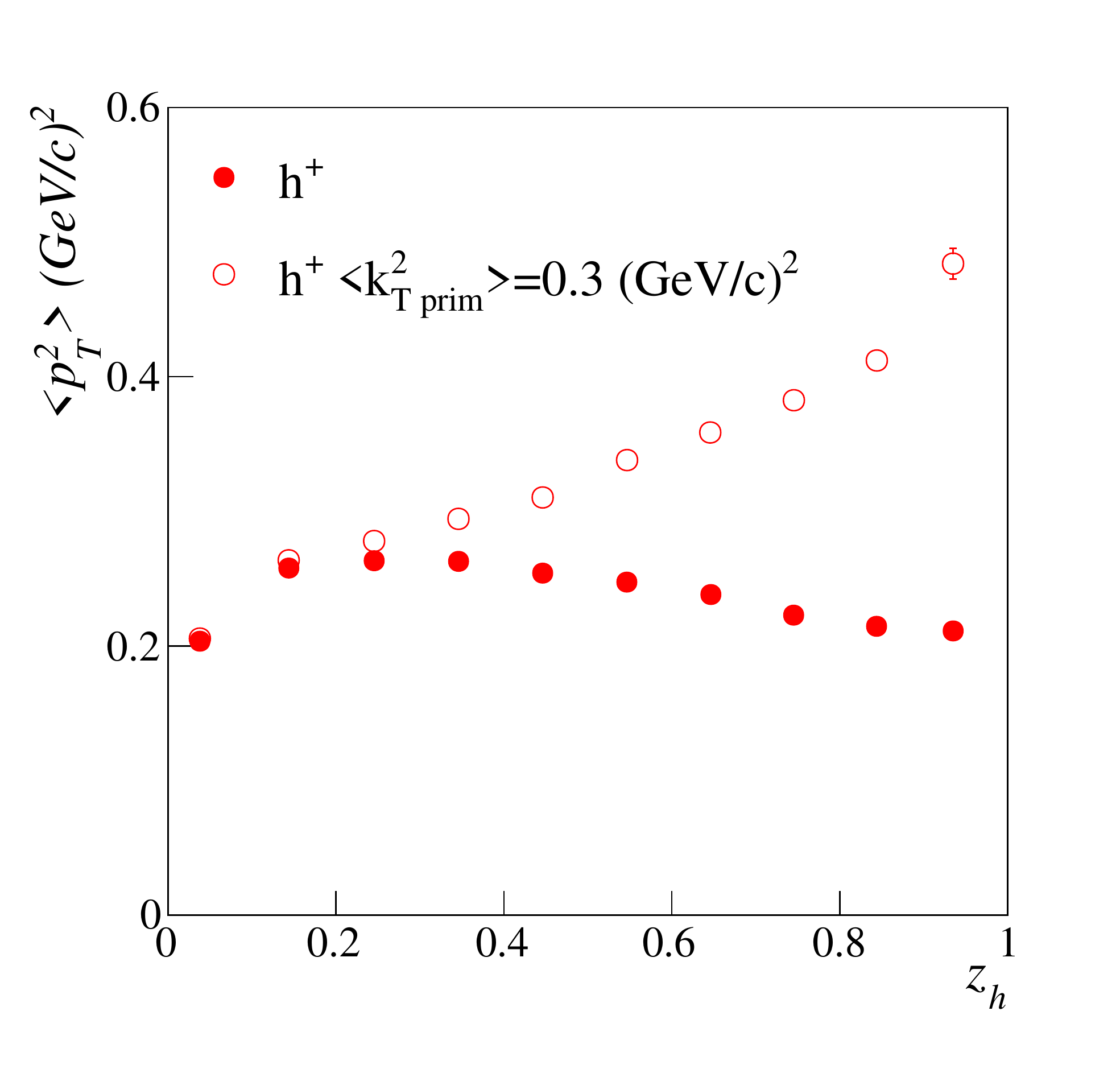}
 \end{minipage}
 \caption{$\langle \ptpt\rangle$ as function of $z_h$ for positive hadrons without (closed points) and with (open points) the primordial transverse momentum.} \label{fig:pt2 zh kt}
 \end{figure}

From Eq. (\ref{eq:shift}) follows at fixed $x_F$
\begin{equation}
\langle \textbf{p}_{\rm T/\textbf{q}_{\gamma}}^2\rangle = x_F^2\langle \textbf{k}_{\rm T prim}^2\rangle +\langle \ptpt\rangle.
\end{equation}
The effect of $\textbf{k}_{\rm T prim}$ is clearly seen in Fig. \ref{fig:pt2 zh kt} showing the $\langle \textbf{p}^2_{\rm T/\textbf{q}_{\gamma}}\rangle$ as function of $z_h$ for positive hadrons when the fragmenting quark has $\langle \textbf{k}^2_{\rm T prim} \rangle =0.3\,(GeV/c)^2$. The large $z_h$ region, where $z_h\simeq x_F$, is more sensitive to the primordial transverse momentum and the effect decays for smaller values of $z_h$. It turns out that the difference between $\langle \ptpt\rangle$ for positive and negative hadrons shown in Fig. \ref{fig:pt2 zh} is somewhat reduced due to the $x_F^2\langle \textbf{k}^2_{\rm T prim}\rangle$ term but still the negative hadrons are produced with larger transverse momenta.
  
\begin{figure*}[tb]\centering
\begin{minipage}{.7\textwidth}
  \includegraphics[width=0.9\textwidth]{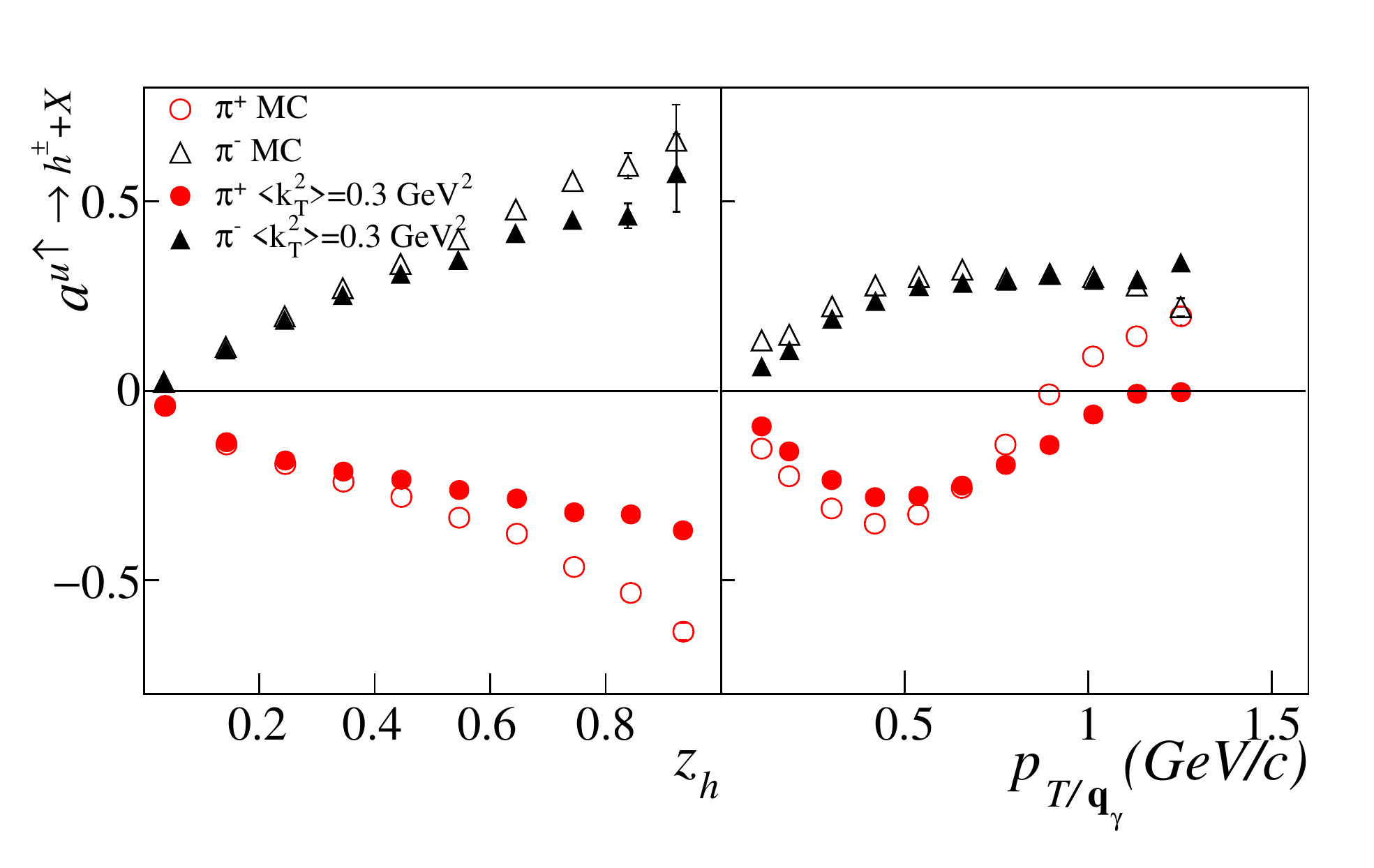}
\end{minipage}%
\caption{\small{Collins analyzing power for charged pions as function of $z_{h}$ (left) and $p_{\rm T/\textbf{q}_{\gamma}}$ (right) for $\langle k_{\rm T}^2\rangle = 0$ (open points) and for $\langle k_{\rm T}^2\rangle = 0.3\,GeV^2/c^2$ (closed points).}}\label{fig:kteffect}
\end{figure*}

In Fig. \ref{fig:kteffect} we show the effect of the primordial transverse momentum on the Collins analyzing power as function of $z_{h}$ (left plot) and $p_{\rm T/\textbf{q}_{\gamma}}$ (right plot) for positive and negative pions. The analysing power for $\langle k_{\rm T prim}^2\rangle=0.3\,(GeV/c)^2$ (full points) is compared to that for vanishing primordial transverse momentum (open points). The reduction of the analysing power is visible at large $z_h$ (left plot) and at low $p_{\rm T/\textbf{q}_{\gamma}}$ (right plot). We note also that the change of sign of the analysing power as function of $p_{\rm T/q_{\gamma}}$ is no more there.
The same effects are also observed for charged kaons.


Table \ref{tab:kteffect} shows the mean values of the single hadron and dihadron analysing powers for charged pions for different values of $\langle \textbf{k}^2_{\rm T prim}\rangle$. At variance with the Collins asymmetry for single hadrons, the asymmetry for pairs of oppositely charged hadrons is not affected by the noise introduced by $\textbf{k}_{\rm T prim}$.

\begin{table}[tb]
\caption{\label{tab:kteffect}\small{Mean value of the analyzing powers shown in Fig.\ref{fig:kteffect} (left) for positive and negative pions with cuts $z_h>0.1$ and $p_{\rm T/\textbf{q}_{\gamma}}>0.1\,(GeV/c)$ have been applied. We show also the mean values of the asymmetry for $\pi^+\pi^-$ pairs with the same cuts.}}
\begin{ruledtabular}
\begin{tabular}{l*{6}{c}r}
$\langle k_{\rm T prim}^2\rangle$         & $\langle a^{u\uparrow\rightarrow \pi^+ +X}\rangle$ & $\langle a^{u\uparrow\rightarrow \pi^- +X}\rangle$  &$\langle a^{u\uparrow\rightarrow \pi^+\pi^- +X}\rangle$ \\
\hline
no $\textbf{k}_{\rm T/prim}$ & $-0.208\pm 0.001$ & $0.188\pm 0.002$  & $-0.276\pm0.002$\\
$0.1\,(GeV/c)^2$ & $-0.197\pm 0.001$ & $0.181\pm 0.002$ &$-0.271\pm0.002$ \\
$0.3\,(GeV/c)^2$ & $-0.183\pm 0.001$ & $0.175\pm 0.002$ &$ -0.269\pm0.002$ \\
$0.5\,(GeV/c)^2$ & $-0.172\pm 0.001$ & $0.169\pm 0.002$ & $-0.266\pm0.002$ \\
\end{tabular}
\end{ruledtabular}
\end{table}


\section{Results on the jet handedness}\label{sec:handedness}
The present model can treat both longitudinal and transverse polarizations at the same time. In particular it can predict jet handedness \cite{Nachtmann,Donoghue,Efremov} which for a particle pair $h_1h_2$ and for a longitudinally polarized quark $q\A$ can be parametrized in the form
\begin{equation}\label{eq:handedness}
\frac{dN_{h_1h_2}}{d^3\textbf{p}_1d^3\textbf{p}_2}\propto 1+a_{\rm JH}^{\vec{q_A}\rightarrow h_1h_2+X}S_{\rm A L}\sin(\phi_2-\phi_1).
\end{equation}
The simplified model of Ref. \cite{DS09} predicts such an effect with an analysing power proportional to $\IM(\mu^2)$. The same factor appears in the present model.

\begin{figure*}[tb]\centering
\begin{minipage}{.7\textwidth}
  \includegraphics[width=0.9\textwidth]{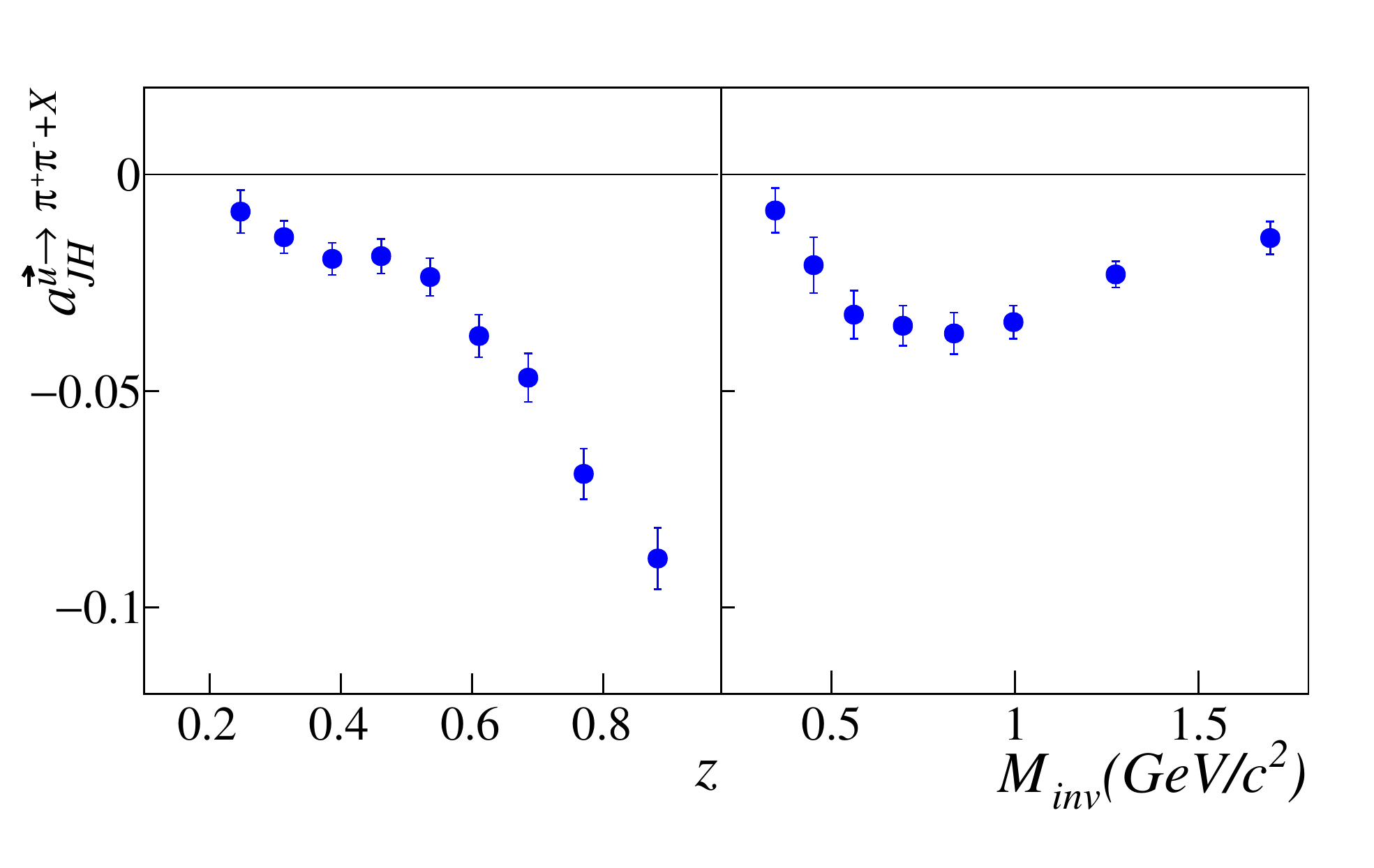}
\end{minipage}%
\caption{\small{The analysing power for the jet handedness effect in the fragmentation $\vec{u}\rightarrow\pi^+\pi^-+X$, defined in Eq. (\ref{eq:handedness}), as function of $z=z_{h_1}+z_{h_2}$ (left plot) and as function of the invariant mass of the pion pair (right plot). The cuts $z_{1(2)}>0.1$ and $p_{1(2)\rm T}>0.1\,GeV/c$ have been applied.}}
\label{fig:handedness}
\end{figure*}

We have made a simulation for $\pi^+\pi^-$ pairs in the jet of an initial longitudinally polarized $u$ quark and calculated the analysing power $a_{\rm JH}^{\vec{u}\rightarrow\pi^+\pi^-+X}$ as $2\langle\sin(\phi_2-\phi_1)\rangle$.
Figure \ref{fig:handedness} shows the dependences of $a_{\rm JH}^{\vec{u}\rightarrow\pi^+\pi^-+X}$ on the invariant mass $M_{inv}$ of the pion pair (left plot) and on the sum of their fractional energies $z_1+z_2$ (right plot). 
While we do not observe a strong dependence on $M_{inv}$, the handedness analysing power increases with $z_1+z_2$. This is expected since at large $z_1+z_2$ both hadrons have nearly fixed ranks (rank $1$ for $\pi^+$ and rank 2 for $\pi^-$).
Comparing Fig. \ref{fig:handedness} with Fig. \ref{fig:zh compass}, where we remind that the Monte Carlo analysing power is scaled by the factor $\lambda_2$, we find for the jet handedness an effect smaller than the dihadron asymmetry by one order of magnitude.
The signal may be improved by imposing further conditions like $z_1>z_2$ or weighting by a power of $p_{1\rm T}$ and $p_{2\rm T}$.

Up to now, attempts to observe jet handedness were not conclusive, see $e.g.$ Ref. \cite{Abe}. Several reasons can explain this failure: 
\begin{itemize}
\item[-] the sign of the asymmetry may vary too much with the charges, the rapidity ordering or the invariant mass of the $h_1-h_2$ pair. 
\item[-] the observable $cos(\phi_2 - \phi_1)$ is very sensitive to a redefinition of the jet axis. It can be easily blurred by experimental uncertainty on jet axis orientation and by gluon radiation. 
\end{itemize}
Like for the Collins effect, the blurring effect can be eliminated by involving one more particle. Indeed, for three particles $h_1$, $h_2$ and $h_3$ of the jet, the pseudoscalar quantity
\begin{equation}
 J = (\textbf{p}_1 \times \textbf{p}_2)  \cdot \textbf{p}_3  = (\textbf{p}_{1,\perp P} \times \textbf{p}_{2,\perp P})  \cdot \textbf{P},
\end{equation}
where $\Pv=\pv_1+\pv_2+\pv_3$, is independent of the jet axis and we may take $\langle J \rangle $ as helicity-sensitive  estimator (the estimator $\langle \rm{sign}(J) \rangle $ was proposed in Ref. \cite{ Efremov}). However it requires the clean measurement of three particle momenta and its amplitude depends on a 6 kinematical variables, $e.g.$, $z_1$, $z_2$, $z_3$, $ |\textbf{p}_{1,\perp P}|$ , $ |\textbf{p}_{2,\perp P}|$  and $ |\textbf{p}_{3,\perp P}|$. 

\section{Conclusions and perspectives}
We have developed a stand alone Monte Carlo code for the simulation of the fragmentation process of a polarized quark ($u$, $d$ or $s$). The theoretical framework is provided by the string fragmentation model where the quark-antiquark pairs in the string cutting points are produced according to the ${}^3P_0$ mechanism. The quark spin is included through spin density matrices and propagated along the decay chain reproducing the $string+{}^3P_0$ mechanism.

With respect to the Lund Symmetric Model, this model requires an additional complex mass parameter whose imaginary part directly affects the single hadron Collins asymmetry. The three free parameters present in the string fragmentation framework and the absolute value of the complex mass have been tuned by comparison with unpolarized experimental SIDIS data.

The analysing powers have been extracted from the simulated events both for the single hadron and for the hadron pairs.
The results of the simulation show a Collins analysing power of opposite sign for oppositely charged mesons. The dependence on the kinematical variables has been investigated, finding a good agreement with experimental data. A clearly different from zero analysing power for hadron pairs of opposite sign in the same jet is also obtained from the same simulated data. The Monte Carlo results are effect compared to BELLE and COMPASS dihadron asymmetries finding again a satisfactory agreement.
Furthermore with the same model we predict also a jet handedness effect in the fragmentation of a longitudinally polarized quark.

Such a model can be a guide to optimize the estimators of quark polarimetry.
An interface of our Monte Carlo program with the PYTHIA event generator is foreseen.
A further improvement of the model is the inclusion of resonances, in particular of vector mesons, and the generation of their hadronic decays.



\begin{thebibliography}{33}%
\makeatletter
\providecommand \@ifxundefined [1]{%
 \@ifx{#1\undefined}
}%
\providecommand \@ifnum [1]{%
 \ifnum #1\expandafter \@firstoftwo
 \else \expandafter \@secondoftwo
 \fi
}%
\providecommand \@ifx [1]{%
 \ifx #1\expandafter \@firstoftwo
 \else \expandafter \@secondoftwo
 \fi
}%
\providecommand \natexlab [1]{#1}%
\providecommand \enquote  [1]{``#1''}%
\providecommand \bibnamefont  [1]{#1}%
\providecommand \bibfnamefont [1]{#1}%
\providecommand \citenamefont [1]{#1}%
\providecommand \href@noop [0]{\@secondoftwo}%
\providecommand \href [0]{\begingroup \@sanitize@url \@href}%
\providecommand \@href[1]{\@@startlink{#1}\@@href}%
\providecommand \@@href[1]{\endgroup#1\@@endlink}%
\providecommand \@sanitize@url [0]{\catcode `\\12\catcode `\$12\catcode
  `\&12\catcode `\#12\catcode `\^12\catcode `\_12\catcode `\%12\relax}%
\providecommand \@@startlink[1]{}%
\providecommand \@@endlink[0]{}%
\providecommand \url  [0]{\begingroup\@sanitize@url \@url }%
\providecommand \@url [1]{\endgroup\@href {#1}{\urlprefix }}%
\providecommand \urlprefix  [0]{URL }%
\providecommand \Eprint [0]{\href }%
\providecommand \doibase [0]{http://dx.doi.org/}%
\providecommand \selectlanguage [0]{\@gobble}%
\providecommand \bibinfo  [0]{\@secondoftwo}%
\providecommand \bibfield  [0]{\@secondoftwo}%
\providecommand \translation [1]{[#1]}%
\providecommand \BibitemOpen [0]{}%
\providecommand \bibitemStop [0]{}%
\providecommand \bibitemNoStop [0]{.\EOS\space}%
\providecommand \EOS [0]{\spacefactor3000\relax}%
\providecommand \BibitemShut  [1]{\csname bibitem#1\endcsname}%
\let\auto@bib@innerbib\@empty
\bibitem [{\citenamefont {Field}\ and\ \citenamefont
  {Feynman}(1977)}]{feynman-fieldw}%
  \BibitemOpen
  \bibfield  {author} {\bibinfo {author} {\bibfnamefont {R.~D.}\ \bibnamefont
  {Field}}\ and\ \bibinfo {author} {\bibfnamefont {R.~P.}\ \bibnamefont
  {Feynman}},\ }\href {\doibase 10.1103/PhysRevD.15.2590} {\bibfield  {journal}
  {\bibinfo  {journal} {Phys. Rev. D}\ }\textbf {\bibinfo {volume} {15}},\
  \bibinfo {pages} {2590} (\bibinfo {year} {1977})}\BibitemShut {NoStop}%
\bibitem [{\citenamefont {Finkelstein}\ and\ \citenamefont
  {Peccei}(1972)}]{Fi-Pe}%
  \BibitemOpen
  \bibfield  {author} {\bibinfo {author} {\bibfnamefont {J.}~\bibnamefont
  {Finkelstein}}\ and\ \bibinfo {author} {\bibfnamefont {R.~D.}\ \bibnamefont
  {Peccei}},\ }\href {\doibase 10.1103/PhysRevD.6.2606} {\bibfield  {journal}
  {\bibinfo  {journal} {Phys. Rev.}\ }\textbf {\bibinfo {volume} {D6}},\
  \bibinfo {pages} {2606} (\bibinfo {year} {1972})}\BibitemShut {NoStop}%
\bibitem [{\citenamefont {Andersson}\ \emph {et~al.}(1983)\citenamefont
  {Andersson}, \citenamefont {Gustafson}, \citenamefont {Ingelman},\ and\
  \citenamefont {Sjostrand}}]{lund}%
  \BibitemOpen
  \bibfield  {author} {\bibinfo {author} {\bibfnamefont {B.}~\bibnamefont
  {Andersson}}, \bibinfo {author} {\bibfnamefont {G.}~\bibnamefont
  {Gustafson}}, \bibinfo {author} {\bibfnamefont {G.}~\bibnamefont {Ingelman}},
  \ and\ \bibinfo {author} {\bibfnamefont {T.}~\bibnamefont {Sjostrand}},\
  }\href {\doibase 10.1016/0370-1573(83)90080-7} {\bibfield  {journal}
  {\bibinfo  {journal} {Phys. Rept.}\ }\textbf {\bibinfo {volume} {97}},\
  \bibinfo {pages} {31} (\bibinfo {year} {1983})}\BibitemShut {NoStop}%
\bibitem [{\citenamefont {Sjostrand}\ \emph {et~al.}(2006)\citenamefont
  {Sjostrand}, \citenamefont {Mrenna},\ and\ \citenamefont {Skands}}]{pythia}%
  \BibitemOpen
  \bibfield  {author} {\bibinfo {author} {\bibfnamefont {T.}~\bibnamefont
  {Sjostrand}}, \bibinfo {author} {\bibfnamefont {S.}~\bibnamefont {Mrenna}}, \
  and\ \bibinfo {author} {\bibfnamefont {P.~Z.}\ \bibnamefont {Skands}},\
  }\href {\doibase 10.1088/1126-6708/2006/05/026} {\bibfield  {journal}
  {\bibinfo  {journal} {JHEP}\ }\textbf {\bibinfo {volume} {05}},\ \bibinfo
  {pages} {026} (\bibinfo {year} {2006})},\ \Eprint
  {http://arxiv.org/abs/hep-ph/0603175} {arXiv:hep-ph/0603175 [hep-ph]}
  \BibitemShut {NoStop}%
\bibitem [{\citenamefont {Artru}(2009)}]{DS09}%
  \BibitemOpen
  \bibfield  {author} {\bibinfo {author} {\bibfnamefont {X.}~\bibnamefont
  {Artru}},\ }in\ \href {http://hal.in2p3.fr/in2p3-00953539} {\emph {\bibinfo
  {booktitle} {{Proc. of XIII Advanced Research Workshop on High Energy Spin
  Physics (DSPIN-09) (Dubna, September 1-5, 2009)}}}},\ \bibinfo {editor}
  {edited by\ \bibinfo {editor} {\bibfnamefont {A.}~\bibnamefont {Efremov}}\
  and\ \bibinfo {editor} {\bibfnamefont {S.}~\bibnamefont {Goloskokov}}}\
  (\bibinfo {address} {Dubna, JINR.},\ \bibinfo {year} {2009})\ p.\ \bibinfo
  {pages} {33.},\ \Eprint {http://arxiv.org/abs/1001.1061} {arXiv:1001.1061
  [hep-ph]} \BibitemShut {NoStop}%
\bibitem [{\citenamefont {Artru}\ and\ \citenamefont {Belghobsi}(2011)}]{DS11}%
  \BibitemOpen
  \bibfield  {author} {\bibinfo {author} {\bibfnamefont {X.}~\bibnamefont
  {Artru}}\ and\ \bibinfo {author} {\bibfnamefont {Z.}~\bibnamefont
  {Belghobsi}},\ }in\ \href {http://hal.in2p3.fr/in2p3-00672599} {\emph
  {\bibinfo {booktitle} {{XIV Advanced Research Workshop on High Spin Physics
  (DSPIN-11)}}}}\ (\bibinfo {address} {Dubna, Russia},\ \bibinfo {year}
  {2011})\BibitemShut {NoStop}%
\bibitem [{\citenamefont {Artru}\ and\ \citenamefont {Belghobsi}(2013)}]{DS13}%
  \BibitemOpen
  \bibfield  {author} {\bibinfo {author} {\bibfnamefont {X.}~\bibnamefont
  {Artru}}\ and\ \bibinfo {author} {\bibfnamefont {Z.}~\bibnamefont
  {Belghobsi}},\ }in\ \href {http://hal.in2p3.fr/in2p3-00953539} {\emph
  {\bibinfo {booktitle} {{XV Advanced Research Workshop on High Energy Spin
  Physics}}}},\ \bibinfo {editor} {edited by\ \bibinfo {editor} {\bibfnamefont
  {A.}~\bibnamefont {Efremov}}\ and\ \bibinfo {editor} {\bibfnamefont
  {S.}~\bibnamefont {Goloskokov}}}\ (\bibinfo {address} {Dubna, Russia},\
  \bibinfo {year} {2013})\ pp.\ \bibinfo {pages} {33--40}\BibitemShut {NoStop}%
\bibitem [{\citenamefont {Artru}\ \emph {et~al.}(1997)\citenamefont {Artru},
  \citenamefont {Czyzewski},\ and\ \citenamefont {Yabuki}}]{XA-JCZ}%
  \BibitemOpen
  \bibfield  {author} {\bibinfo {author} {\bibfnamefont {X.}~\bibnamefont
  {Artru}}, \bibinfo {author} {\bibfnamefont {J.}~\bibnamefont {Czyzewski}}, \
  and\ \bibinfo {author} {\bibfnamefont {H.}~\bibnamefont {Yabuki}},\ }\href
  {\doibase 10.1007/s002880050342} {\bibfield  {journal} {\bibinfo  {journal}
  {Z. Phys.}\ }\textbf {\bibinfo {volume} {C73}},\ \bibinfo {pages} {527}
  (\bibinfo {year} {1997})},\ \Eprint {http://arxiv.org/abs/hep-ph/9508239}
  {arXiv:hep-ph/9508239 [hep-ph]} \BibitemShut {NoStop}%
\bibitem [{\citenamefont {Kerbizi}(2016)}]{albi-tesi}%
  \BibitemOpen
  \bibfield  {author} {\bibinfo {author} {\bibfnamefont {A.}~\bibnamefont
  {Kerbizi}},\ }\emph {\bibinfo {title} {Study of the fragmentation process of
  transversely polarized quarks}},\ \href@noop {} {Master's thesis},\ \bibinfo
  {school} {University of Trieste}, \bibinfo {address}
  {$http://www.infn.it/thesis/thesis_dettaglio.php?tid=11076$} (\bibinfo {year}
  {2016})\BibitemShut {NoStop}%
\bibitem [{\citenamefont {Kerbizi}\ \emph {et~al.}(2017)\citenamefont
  {Kerbizi}, \citenamefont {Artru}, \citenamefont {Belghobsi}, \citenamefont
  {Bradamante}, \citenamefont {Martin},\ and\ \citenamefont {Salah}}]{spin16}%
  \BibitemOpen
  \bibfield  {author} {\bibinfo {author} {\bibfnamefont {A.}~\bibnamefont
  {Kerbizi}}, \bibinfo {author} {\bibfnamefont {X.}~\bibnamefont {Artru}},
  \bibinfo {author} {\bibfnamefont {Z.}~\bibnamefont {Belghobsi}}, \bibinfo
  {author} {\bibfnamefont {F.}~\bibnamefont {Bradamante}}, \bibinfo {author}
  {\bibfnamefont {A.}~\bibnamefont {Martin}}, \ and\ \bibinfo {author}
  {\bibfnamefont {E.~R.}\ \bibnamefont {Salah}},\ }in\ \href
  {http://inspirehep.net/record/1511304/files/arXiv:1701.08543.pdf} {\emph
  {\bibinfo {booktitle} {{22nd International Symposium on Spin Physics (SPIN
  2016) Urbana, IL, USA, September 25-30, 2016}}}}\ (\bibinfo {year} {2017})\
  \Eprint {http://arxiv.org/abs/1701.08543} {arXiv:1701.08543 [hep-ph]}
  \BibitemShut {NoStop}%
\bibitem [{\citenamefont {Matevosyan}\ \emph {et~al.}(2017)\citenamefont
  {Matevosyan}, \citenamefont {Kotzinian},\ and\ \citenamefont
  {Thomas}}]{Mate}%
  \BibitemOpen
  \bibfield  {author} {\bibinfo {author} {\bibfnamefont {H.~H.}\ \bibnamefont
  {Matevosyan}}, \bibinfo {author} {\bibfnamefont {A.}~\bibnamefont
  {Kotzinian}}, \ and\ \bibinfo {author} {\bibfnamefont {A.~W.}\ \bibnamefont
  {Thomas}},\ }\href {\doibase 10.1103/PhysRevD.95.014021} {\bibfield
  {journal} {\bibinfo  {journal} {Phys. Rev.}\ }\textbf {\bibinfo {volume}
  {D95}},\ \bibinfo {pages} {014021} (\bibinfo {year} {2017})},\ \Eprint
  {http://arxiv.org/abs/1610.05624} {arXiv:1610.05624 [hep-ph]} \BibitemShut
  {NoStop}%
\bibitem [{\citenamefont {Artru}\ and\ \citenamefont
  {Mennessier}(1974)}]{A-Men}%
  \BibitemOpen
  \bibfield  {author} {\bibinfo {author} {\bibfnamefont {X.}~\bibnamefont
  {Artru}}\ and\ \bibinfo {author} {\bibfnamefont {G.}~\bibnamefont
  {Mennessier}},\ }\href {\doibase 10.1016/0550-3213(74)90360-5} {\bibfield
  {journal} {\bibinfo  {journal} {Nucl. Phys.}\ }\textbf {\bibinfo {volume}
  {B70}},\ \bibinfo {pages} {93} (\bibinfo {year} {1974})}\BibitemShut
  {NoStop}%
\bibitem [{\citenamefont {Artru}(1984)}]{XA1984}%
  \BibitemOpen
  \bibfield  {author} {\bibinfo {author} {\bibfnamefont {X.}~\bibnamefont
  {Artru}},\ }\href {\doibase 10.1007/BF01572545} {\bibfield  {journal}
  {\bibinfo  {journal} {Z. Phys.}\ }\textbf {\bibinfo {volume} {C26}},\
  \bibinfo {pages} {83} (\bibinfo {year} {1984})}\BibitemShut {NoStop}%
\bibitem [{\citenamefont {Andersson}\ and\ \citenamefont
  {Hofmann}(1986)}]{BA-Hof}%
  \BibitemOpen
  \bibfield  {author} {\bibinfo {author} {\bibfnamefont {B.}~\bibnamefont
  {Andersson}}\ and\ \bibinfo {author} {\bibfnamefont {W.}~\bibnamefont
  {Hofmann}},\ }\href {\doibase 10.1016/0370-2693(86)90373-4} {\bibfield
  {journal} {\bibinfo  {journal} {Phys. Lett.}\ }\textbf {\bibinfo {volume}
  {169B}},\ \bibinfo {pages} {364} (\bibinfo {year} {1986})}\BibitemShut
  {NoStop}%
\bibitem [{\citenamefont {Artru}\ and\ \citenamefont {Bowler}(1988)}]{QYD}%
  \BibitemOpen
  \bibfield  {author} {\bibinfo {author} {\bibfnamefont {X.}~\bibnamefont
  {Artru}}\ and\ \bibinfo {author} {\bibfnamefont {M.~G.}\ \bibnamefont
  {Bowler}},\ }\href {\doibase 10.1007/BF01579915} {\bibfield  {journal}
  {\bibinfo  {journal} {Z. Phys.}\ }\textbf {\bibinfo {volume} {C37}},\
  \bibinfo {pages} {293} (\bibinfo {year} {1988})}\BibitemShut {NoStop}%
\bibitem [{\citenamefont {Collins}\ and\ \citenamefont
  {Rogers}(2018)}]{collins-rogers}%
  \BibitemOpen
  \bibfield  {author} {\bibinfo {author} {\bibfnamefont {J.}~\bibnamefont
  {Collins}}\ and\ \bibinfo {author} {\bibfnamefont {T.~C.}\ \bibnamefont
  {Rogers}},\ }\href@noop {} {\  (\bibinfo {year} {2018})},\ \Eprint
  {http://arxiv.org/abs/1801.02704} {arXiv:1801.02704 [hep-ph]} \BibitemShut
  {NoStop}%
\bibitem [{\citenamefont {Artru}\ \emph {et~al.}(2016)\citenamefont {Artru},
  \citenamefont {Belghobsi},\ and\ \citenamefont {Redouane-Salah}}]{LUG}%
  \BibitemOpen
  \bibfield  {author} {\bibinfo {author} {\bibfnamefont {X.}~\bibnamefont
  {Artru}}, \bibinfo {author} {\bibfnamefont {Z.}~\bibnamefont {Belghobsi}}, \
  and\ \bibinfo {author} {\bibfnamefont {E.}~\bibnamefont {Redouane-Salah}},\
  }\href {\doibase 10.1103/PhysRevD.94.034034} {\bibfield  {journal} {\bibinfo
  {journal} {Phys. Rev.}\ }\textbf {\bibinfo {volume} {D94}},\ \bibinfo {pages}
  {034034} (\bibinfo {year} {2016})},\ \Eprint
  {http://arxiv.org/abs/1607.07106} {arXiv:1607.07106 [hep-ph]} \BibitemShut
  {NoStop}%
\bibitem [{\citenamefont {Aghasyan}\ \emph {et~al.}(2017)\citenamefont
  {Aghasyan} \emph {et~al.}}]{compass-pt2}%
  \BibitemOpen
  \bibfield  {author} {\bibinfo {author} {\bibfnamefont {M.}~\bibnamefont
  {Aghasyan}} \emph {et~al.} (\bibinfo {collaboration} {COMPASS}),\ }\href@noop
  {} {\  (\bibinfo {year} {2017})},\ \Eprint {http://arxiv.org/abs/1709.07374}
  {arXiv:1709.07374 [hep-ex]} \BibitemShut {NoStop}%
\bibitem [{\citenamefont {Kniehl}\ \emph {et~al.}(2001)\citenamefont {Kniehl},
  \citenamefont {Kramer},\ and\ \citenamefont
  {P{\"o}tter}}]{kniehl2001testing}%
  \BibitemOpen
  \bibfield  {author} {\bibinfo {author} {\bibfnamefont {B.~A.}\ \bibnamefont
  {Kniehl}}, \bibinfo {author} {\bibfnamefont {G.}~\bibnamefont {Kramer}}, \
  and\ \bibinfo {author} {\bibfnamefont {B.}~\bibnamefont {P{\"o}tter}},\
  }\href@noop {} {\bibfield  {journal} {\bibinfo  {journal} {Nuclear Physics
  B}\ }\textbf {\bibinfo {volume} {597}},\ \bibinfo {pages} {337} (\bibinfo
  {year} {2001})}\BibitemShut {NoStop}%
\bibitem [{\citenamefont {Airapetian}\ \emph {et~al.}(2013)\citenamefont
  {Airapetian} \emph {et~al.}}]{hermes}%
  \BibitemOpen
  \bibfield  {author} {\bibinfo {author} {\bibfnamefont {A.}~\bibnamefont
  {Airapetian}} \emph {et~al.} (\bibinfo {collaboration} {HERMES}),\ }\href
  {\doibase 10.1103/PhysRevD.87.012010} {\bibfield  {journal} {\bibinfo
  {journal} {Phys. Rev.}\ }\textbf {\bibinfo {volume} {D87}},\ \bibinfo {pages}
  {012010} (\bibinfo {year} {2013})},\ \Eprint {http://arxiv.org/abs/1204.4161}
  {arXiv:1204.4161 [hep-ex]} \BibitemShut {NoStop}%
\bibitem [{\citenamefont {Qian}\ \emph {et~al.}(2011)\citenamefont {Qian} \emph
  {et~al.}}]{jlab}%
  \BibitemOpen
  \bibfield  {author} {\bibinfo {author} {\bibfnamefont {X.}~\bibnamefont
  {Qian}} \emph {et~al.} (\bibinfo {collaboration} {Jefferson Lab Hall A}),\
  }\href {\doibase 10.1103/PhysRevLett.107.072003} {\bibfield  {journal}
  {\bibinfo  {journal} {Phys. Rev. Lett.}\ }\textbf {\bibinfo {volume} {107}},\
  \bibinfo {pages} {072003} (\bibinfo {year} {2011})},\ \Eprint
  {http://arxiv.org/abs/1106.0363} {arXiv:1106.0363 [nucl-ex]} \BibitemShut
  {NoStop}%
\bibitem [{\citenamefont {Lees}\ \emph {et~al.}(2014)\citenamefont {Lees} \emph
  {et~al.}}]{babar}%
  \BibitemOpen
  \bibfield  {author} {\bibinfo {author} {\bibfnamefont {J.~P.}\ \bibnamefont
  {Lees}} \emph {et~al.} (\bibinfo {collaboration} {BaBar}),\ }\href {\doibase
  10.1103/PhysRevD.90.052003} {\bibfield  {journal} {\bibinfo  {journal} {Phys.
  Rev.}\ }\textbf {\bibinfo {volume} {D90}},\ \bibinfo {pages} {052003}
  (\bibinfo {year} {2014})},\ \Eprint {http://arxiv.org/abs/1309.5278}
  {arXiv:1309.5278 [hep-ex]} \BibitemShut {NoStop}%
\bibitem [{\citenamefont {Ablikim}\ \emph {et~al.}(2016)\citenamefont {Ablikim}
  \emph {et~al.}}]{besIII}%
  \BibitemOpen
  \bibfield  {author} {\bibinfo {author} {\bibfnamefont {M.}~\bibnamefont
  {Ablikim}} \emph {et~al.} (\bibinfo {collaboration} {BESIII}),\ }\href
  {\doibase 10.1103/PhysRevLett.116.042001} {\bibfield  {journal} {\bibinfo
  {journal} {Phys. Rev. Lett.}\ }\textbf {\bibinfo {volume} {116}},\ \bibinfo
  {pages} {042001} (\bibinfo {year} {2016})},\ \Eprint
  {http://arxiv.org/abs/1507.06824} {arXiv:1507.06824 [hep-ex]} \BibitemShut
  {NoStop}%
\bibitem [{\citenamefont {Collins}(1993)}]{FFcollins}%
  \BibitemOpen
  \bibfield  {author} {\bibinfo {author} {\bibfnamefont {J.~C.}\ \bibnamefont
  {Collins}},\ }\href {\doibase 10.1016/0550-3213(93)90262-N} {\bibfield
  {journal} {\bibinfo  {journal} {Nucl. Phys.}\ }\textbf {\bibinfo {volume}
  {B396}},\ \bibinfo {pages} {161} (\bibinfo {year} {1993})},\ \Eprint
  {http://arxiv.org/abs/hep-ph/9208213} {arXiv:hep-ph/9208213 [hep-ph]}
  \BibitemShut {NoStop}%
\bibitem [{\citenamefont {Martin}\ \emph {et~al.}(2015)\citenamefont {Martin},
  \citenamefont {Bradamante},\ and\ \citenamefont {Barone}}]{M.B.B}%
  \BibitemOpen
  \bibfield  {author} {\bibinfo {author} {\bibfnamefont {A.}~\bibnamefont
  {Martin}}, \bibinfo {author} {\bibfnamefont {F.}~\bibnamefont {Bradamante}},
  \ and\ \bibinfo {author} {\bibfnamefont {V.}~\bibnamefont {Barone}},\ }\href
  {\doibase 10.1103/PhysRevD.91.014034} {\bibfield  {journal} {\bibinfo
  {journal} {Phys. Rev.}\ }\textbf {\bibinfo {volume} {D91}},\ \bibinfo {pages}
  {014034} (\bibinfo {year} {2015})},\ \Eprint {http://arxiv.org/abs/1412.5946}
  {arXiv:1412.5946 [hep-ph]} \BibitemShut {NoStop}%
\bibitem [{\citenamefont {Adolph}\ \emph {et~al.}(2015)\citenamefont {Adolph}
  \emph {et~al.}}]{compassplb}%
  \BibitemOpen
  \bibfield  {author} {\bibinfo {author} {\bibfnamefont {C.}~\bibnamefont
  {Adolph}} \emph {et~al.} (\bibinfo {collaboration} {COMPASS}),\ }\href
  {\doibase 10.1016/j.physletb.2015.03.056} {\bibfield  {journal} {\bibinfo
  {journal} {Phys. Lett.}\ }\textbf {\bibinfo {volume} {B744}},\ \bibinfo
  {pages} {250} (\bibinfo {year} {2015})},\ \Eprint
  {http://arxiv.org/abs/1408.4405} {arXiv:1408.4405 [hep-ex]} \BibitemShut
  {NoStop}%
\bibitem [{\citenamefont {Adolph}\ \emph {et~al.}(2016)\citenamefont {Adolph}
  \emph {et~al.}}]{interplay}%
  \BibitemOpen
  \bibfield  {author} {\bibinfo {author} {\bibfnamefont {C.}~\bibnamefont
  {Adolph}} \emph {et~al.} (\bibinfo {collaboration} {COMPASS}),\ }\href
  {\doibase 10.1016/j.physletb.2015.12.042} {\bibfield  {journal} {\bibinfo
  {journal} {Phys. Lett.}\ }\textbf {\bibinfo {volume} {B753}},\ \bibinfo
  {pages} {406} (\bibinfo {year} {2016})},\ \Eprint
  {http://arxiv.org/abs/1507.07593} {arXiv:1507.07593 [hep-ex]} \BibitemShut
  {NoStop}%
\bibitem [{\citenamefont {Vossen}\ \emph {et~al.}(2011)\citenamefont {Vossen}
  \emph {et~al.}}]{belle}%
  \BibitemOpen
  \bibfield  {author} {\bibinfo {author} {\bibfnamefont {A.}~\bibnamefont
  {Vossen}} \emph {et~al.} (\bibinfo {collaboration} {Belle}),\ }\href
  {\doibase 10.1103/PhysRevLett.107.072004} {\bibfield  {journal} {\bibinfo
  {journal} {Phys. Rev. Lett.}\ }\textbf {\bibinfo {volume} {107}},\ \bibinfo
  {pages} {072004} (\bibinfo {year} {2011})},\ \Eprint
  {http://arxiv.org/abs/1104.2425} {arXiv:1104.2425 [hep-ex]} \BibitemShut
  {NoStop}%
\bibitem [{\citenamefont {Adolph}\ \emph {et~al.}(2014)\citenamefont {Adolph}
  \emph {et~al.}}]{compass-dihadron}%
  \BibitemOpen
  \bibfield  {author} {\bibinfo {author} {\bibfnamefont {C.}~\bibnamefont
  {Adolph}} \emph {et~al.} (\bibinfo {collaboration} {COMPASS}),\ }\href
  {\doibase 10.1016/j.physletb.2014.06.080} {\bibfield  {journal} {\bibinfo
  {journal} {Phys. Lett.}\ }\textbf {\bibinfo {volume} {B736}},\ \bibinfo
  {pages} {124} (\bibinfo {year} {2014})},\ \Eprint
  {http://arxiv.org/abs/1401.7873} {arXiv:1401.7873 [hep-ex]} \BibitemShut
  {NoStop}%
\bibitem [{\citenamefont {Nachtmann}(1977)}]{Nachtmann}%
  \BibitemOpen
  \bibfield  {author} {\bibinfo {author} {\bibfnamefont {O.}~\bibnamefont
  {Nachtmann}},\ }\href {\doibase 10.1016/0550-3213(77)90217-6} {\bibfield
  {journal} {\bibinfo  {journal} {Nucl. Phys.}\ }\textbf {\bibinfo {volume}
  {B127}},\ \bibinfo {pages} {314} (\bibinfo {year} {1977})}\BibitemShut
  {NoStop}%
\bibitem [{\citenamefont {Donoghue}(1979)}]{Donoghue}%
  \BibitemOpen
  \bibfield  {author} {\bibinfo {author} {\bibfnamefont {J.~F.}\ \bibnamefont
  {Donoghue}},\ }\href {\doibase 10.1103/PhysRevD.19.2806} {\bibfield
  {journal} {\bibinfo  {journal} {Phys. Rev.}\ }\textbf {\bibinfo {volume}
  {D19}},\ \bibinfo {pages} {2806} (\bibinfo {year} {1979})}\BibitemShut
  {NoStop}%
\bibitem [{\citenamefont {Efremov}\ \emph {et~al.}(1992)\citenamefont
  {Efremov}, \citenamefont {Mankiewicz},\ and\ \citenamefont
  {Tornqvist}}]{Efremov}%
  \BibitemOpen
  \bibfield  {author} {\bibinfo {author} {\bibfnamefont {A.~V.}\ \bibnamefont
  {Efremov}}, \bibinfo {author} {\bibfnamefont {L.}~\bibnamefont {Mankiewicz}},
  \ and\ \bibinfo {author} {\bibfnamefont {N.~A.}\ \bibnamefont {Tornqvist}},\
  }\href {\doibase 10.1016/0370-2693(92)90451-9} {\bibfield  {journal}
  {\bibinfo  {journal} {Phys. Lett.}\ }\textbf {\bibinfo {volume} {B284}},\
  \bibinfo {pages} {394} (\bibinfo {year} {1992})}\BibitemShut {NoStop}%
\bibitem [{\citenamefont {Abe}\ \emph {et~al.}(1995)\citenamefont {Abe} \emph
  {et~al.}}]{Abe}%
  \BibitemOpen
  \bibfield  {author} {\bibinfo {author} {\bibfnamefont {K.}~\bibnamefont
  {Abe}} \emph {et~al.} (\bibinfo {collaboration} {SLD}),\ }\href {\doibase
  10.1103/PhysRevLett.74.1512} {\bibfield  {journal} {\bibinfo  {journal}
  {Phys. Rev. Lett.}\ }\textbf {\bibinfo {volume} {74}},\ \bibinfo {pages}
  {1512} (\bibinfo {year} {1995})},\ \Eprint
  {http://arxiv.org/abs/hep-ex/9501006} {arXiv:hep-ex/9501006 [hep-ex]}
  \BibitemShut {NoStop}%
\end{thebibliography}

%

\end{document}